\documentclass[11pt,letterpaper]{article}

\usepackage{graphicx}
\usepackage{geometry}
\usepackage{setspace}
\usepackage{amsfonts,amsmath,amssymb,amsthm} 
\usepackage{mathtools}
\usepackage{bm}

\usepackage[dvipsnames]{xcolor}

\usepackage{booktabs}
\usepackage{multicol}
\usepackage{multirow}
\usepackage{makecell}
\usepackage{hhline}

\usepackage[ruled,linesnumbered,nofillcomment]{algorithm2e}
\usepackage[hyphens]{url}

\usepackage{hyperref}
\usepackage{cleveref}
\hypersetup{
    colorlinks=true,
    linkcolor=Blue,
    citecolor=Mahogany,
    urlcolor=PineGreen
}
\usepackage{crossreftools}
\usepackage{tikz}
\usepackage{pgfplots}
\pgfplotsset{compat=newest}
\usetikzlibrary{shapes, shapes.symbols, backgrounds, matrix, calc, arrows, math, arrows.meta, decorations.pathreplacing, decorations.markings, patterns}

\usepackage{natbib}

\usepackage{extarrows}

\theoremstyle{plain}
\newtheorem{theorem}{Theorem}[section]
\newtheorem{lemma}[theorem]{Lemma}
\newtheorem{proposition}[theorem]{Proposition}
\newtheorem{corollary}[theorem]{Corollary}
\newtheorem{assumption}{Assumption}[section]

\theoremstyle{definition}

\theoremstyle{remark}

\crefname{assumption}{assumption}{assumptions}
\Crefname{assumption}{Assumption}{Assumptions}

\newcommand{\RomanNum}[1]{\uppercase\expandafter{\romannumeral #1}}

\newcommand\algName{OGD-CB }
\newcommand\algoName{OGD-CB }
\newcommand\algNameNoSpace{OGD-CB}

\newcommand\diffe{{\,\mathrm d}}

\newcommand\opt{{\rm OPT}}
\newcommand\OPT{{\rm OPT}}

\newcommand\OPTstrong{{\rm OPT}_{\rm S}}

\newcommand\vmax{\bar{v}}

\newcommand\totalRevenueGamma{U^{\beta}(\bm v, \bm p, \bm \gamma)}
\newcommand\totalRevenueGeneral{U^{\beta}(\bm v, \bm p)}
\newcommand\totalRevenueNoV{U^{\beta}(\bm p, \bm \gamma)}
\newcommand\totalRevenueThrottling{U^{\rm T}(\bm v, \bm p, \bm \gamma)}
\newcommand\totalRevenuePacing{U^{\rm P}(\bm v, \bm p)}
\newcommand\totalRevenueHindsight{U_{\rm H}(\bm v, \bm p)}

\newcommand\regretGamma{Reg^\beta (\bm v, \bm p, \bm \gamma)}
\newcommand\regret{Reg^\beta (\bm v, \bm p)}
\newcommand\regretNoV{Reg^\beta (\bm p, \bm \gamma)}

\newcommand\discountedRegretGamma{Rem^\beta_\mu (\bm v, \bm p, \bm \gamma)}
\newcommand\discountedRegret{Rem^\beta_\mu (\bm v, \bm p)}

\newcommand\OPTthreshold{\lambda^*}
\newcommand\OPTstrongThreshold{\lambda_{\rm S}^*}

\newcommand\estG{\widehat{G}}
\newcommand\estr{\widetilde{r}}
\newcommand\estc{\widetilde{c}}

\newcommand\Kappa{{\rm K}}

\newcommand\constantFull{C^{\rm F}}
\newcommand\constantPartial{C^{\rm P}}
\newcommand\constantFrequency{C_e}
\newcommand\constantLowerBound{C_l}

\allowdisplaybreaks

\title{Dynamic Budget Throttling in Repeated Second-Price Auctions}

\author{
Zhaohua Chen\thanks{These authors contributed equally to this work, and follow a lexicography order. } 
\thanks{CFCS, School of Computer Science, Peking University. Email: \texttt{\{chenzhaohua, charlie, xiaotie\}@pku.edu.cn}}
\and
Chang Wang\footnotemark[1]
\thanks{Northwestern University. Email: \texttt{wc@u.northwestern.edu}}
\and
Qian Wang\footnotemark[1] 
\footnotemark[2] 
\and
Yuqi Pan\thanks{School of Electronics Engineering and Computer Science, Peking University. Email: \texttt{pyq0419@stu.pku.edu.cn}}
\and
Zhuming Shi\thanks{Stony Brook University. Email: \texttt{zhuming.shi@stonybrook.edu}}
\and 
Zheng Cai\thanks{Tencent Technology (Shenzhen) Co., Ltd. Email: \texttt{\{zhengcai, rickyren, zhihuazhu\}@tencent.com}}
\and
Yukun Ren\footnotemark[6]
\and 
Zhihua Zhu\footnotemark[6]
\and
Xiaotie Deng\footnotemark[2]
\thanks{CMAR, Institute for Artificial Intelligence, Peking University. } 
}

\begin{document}

\maketitle

\thispagestyle{empty}
\begin{abstract}
In today's online advertising markets, a crucial requirement for an advertiser is to control her total expenditure within a time horizon under some budget. 
Among various budget control methods, throttling has emerged as a popular choice, managing an advertiser's total expenditure by selecting only a subset of auctions to participate in.
This paper provides a theoretical panorama of a single advertiser's dynamic budget throttling process in repeated second-price auctions.
We first establish a lower bound on the regret and an upper bound on the asymptotic competitive ratio for any throttling algorithm, respectively, when the advertiser's values are stochastic and adversarial. 
Regarding the algorithmic side, we propose the \algName algorithm, which guarantees a near-optimal expected regret with stochastic values. 
On the other hand, when values are adversarial, we prove that this algorithm also reaches the upper bound on the asymptotic competitive ratio. 
We further compare throttling with pacing, another widely adopted budget control method, in repeated second-price auctions. 
In the stochastic case, we demonstrate that pacing is generally superior to throttling for the advertiser, supporting the well-known result that pacing is asymptotically optimal in this scenario. 
However, in the adversarial case, we give an exciting result indicating that throttling is also an asymptotically optimal dynamic bidding strategy. 
Our results bridge the gaps in theoretical research of throttling in repeated auctions and comprehensively reveal the ability of this popular budget-smoothing strategy. 
\end{abstract}

\newpage
\setcounter{page}{1}
\section{Introduction}

In recent years, the online advertising market has experienced significant growth, driven by the rise of new social media platforms such as short videos. 
When a user submits an ad query to the market, an auction is held among all interested advertisers, and the winner is awarded the opportunity to display their ad. 
Owing to the vast market volume, it is common for advertisers to set a budget to regulate their expenditure over a specified period. 
Correspondingly, advertising platforms provide advertisers with various budget control methods to choose from.

This work studies one of these methods, called \emph{throttling} (a.k.a. probabilistic pacing), which is widely adopted by major advertising platforms including Facebook \citep{facebook-throttling}, Google \citep{DBLP:conf/wsdm/KarandeMS13}, LinkedIn \citep{DBLP:conf/kdd/AgarwalGWY14} and Yahoo! \citep{DBLP:conf/kdd/XuLLQL15}. 
Under this method, an advertiser's accumulated payment is controlled by being excluded from a fraction of auctions throughout the entire period (e.g., a day, a week, or a month). 
Compared to other budget management strategies \citep{DBLP:journals/ior/BalseiroKMM21}, such as pacing (a.k.a. bid-shading), an essential feature of throttling is that an advertiser's bid is never altered in any auction instance. 
As revealed in the literature \citep{DBLP:conf/wsdm/KarandeMS13,DBLP:conf/wine/ChenKK21}, this feature attracts a large number of advertisers to use the throttling strategy due to two primary reasons. 
(a) Some advertisers do not permit the platform to modify their bids, forcing the platform to exclude them from some auctions to control their budget, i.e., adopting the throttling strategy. 
(b) Strategies that modify bids, such as pacing, may allocate an advertiser to those auctions where she is superior to other advertisers, and could be detrimental for the advertiser to exploit other parts of the market. 
In contrast, with unmodified bids competing, the throttling strategy provides advertisers with a more straightforward and unbiased observation of various users, enabling them to gain a clearer understanding of their competitiveness in different market sectors. 
For instance, under pacing, a small budget on some market sectors to be explored would lead to a small pacing multiplier on the advertiser's bid, which would cause the advertiser to win nothing in this sector, considering other advertisers who are superior. In comparison, under throttling, even with a small budget, the advertiser has a chance to have a competitive bid (as the bid is not modified) and win in some auctions, thus helping the advertiser to explore the new market sector with a small cost. 
These phenomena illustrate the importance of throttling as a popular budget control method. 

The throttling strategy has been extensively explored in the literature, primarily from an empirical perspective evaluating the performance of specific algorithms  \citep{DBLP:conf/kdd/AgarwalGWY14,DBLP:conf/wsdm/KarandeMS13,DBLP:conf/kdd/XuLLQL15}. 
Some work has also considered the theoretical equilibrium problem when multiple buyers simultaneously adopt the throttling strategy \citep{DBLP:journals/ior/BalseiroKMM21,DBLP:conf/wine/ChenKK21}. 
However, there is currently no literature that takes the perspective of the advertiser and concentrates on how to \emph{theoretically} optimize their accumulated revenue through a throttling strategy in repeated auctions. 
The aim of this paper is to address this problem and design dynamic throttling algorithms that achieve good performance in various input models. 
In practice, such a throttling strategy is implemented through an auto-bidding service that receives the advertiser's values and makes binary choices on behalf of the advertiser in each auction.
However, to simplify the terminology and description, this work shifts the throttling control to the advertiser as a subjective strategy that takes objective values as inputs, which is equivalent to the real-world scenario. 


\subsection{Our Contributions and Techniques}

This work gives a theoretical panorama of an advertiser's dynamic throttling strategy in repeated second-price auctions. 
Specifically, our contributions are three-wise along the following lines. 

\paragraph{Formalization, bounds, and impossibility results.} 
We formalize the problem of dynamic throttling in repeated second-price auctions from the perspective of a single advertiser in two value input models; namely, where private values $\bm v$ are stochastic and adversarial, respectively. 
To model other bidders' bids, we assume the highest competing bids $\bm p$ to be stochastic and following an unknown i.i.d. distribution. 
When $\bm v$ are stochastic, we measure the performance of a throttling strategy by considering its regret and establish an $\Omega(\sqrt{T})$ lower bound on the expected regret (\Cref{thm:lower-bound-stochastic}). 
On the other hand, when $\bm v$ are adversarial, we measure the performance of a throttling strategy by considering its asymptotic competitive ratio and demonstrate that any throttling strategy's asymptotic competitive ratio cannot exceed the advertiser's regularized average budget (\Cref{thm:upper-bound-adversarial}), i.e., the average budget divided by the maximum value. 
To make the last piece of the puzzle, we also consider the circumstance when the highest competing bid $\bm p$ is adversarial and show for this case that any throttling strategy can behave arbitrarily badly even when $\bm v$ is fixed in each round (\Cref{thm:impossibility-result}). 
Therefore, the assumption that $\bm p$ is stochastic would be necessary from the theoretical perspective. 

Following the above bounds results, we propose an \algName algorithm that is oblivious to the value input model. 
The main feature of our algorithm is to combine the online optimization method and the distribution estimation technique. 
For the latter part, notice that though we suppose the stochastic behavior of $\bm p$, its distribution in real life is intricately correlated with multiple factors, including the market size, the preference of competing advertisers, and their active time in the market. 
Consequently, it is almost impossible for a single advertiser to know this distribution \emph{ex-ante}. 
To settle the process of learning such distribution, we consider two information structures in order, with the increasing difficulty of getting a sample. 
With full information feedback, the advertiser could obtain the highest competing bid in each round. 
However, such an assumption could be too strong for practice. 
Therefore, we further consider the partial information feedback structure, in which the advertiser can acquire the highest competing bid only when she participates in an auction.
We mention that the partial feedback model is realizable in the industry, especially for auto-bidding services, if the platform announces the highest bid to all participants after an auction. 

\begin{table}[!t]
    \renewcommand{\arraystretch}{1.25}
\setcellgapes{2.5pt}
\makegapedcells

\centering
\begin{tabular}{|p{12em}<{\centering}||*{4}{p{5em}<{\centering}|}}
    \hline
    \textbf{Value Input Model} & \multicolumn{2}{c|}{Stochastic} & \multicolumn{2}{c|}{Adversarial} \\
    \hline
    \textbf{Information Structure} & Full & Partial & Full & Partial \\
    \hhline{|=#==|==|}
    \textbf{Bounds} & \multicolumn{2}{c|}{\makecell{$\Omega(\sqrt{T})$ regret \\ (\Cref{thm:lower-bound-stochastic}) }} & \multicolumn{2}{c|}{\makecell{$(\rho / \vmax)$-asymptotic \\ competitive ratio \\ (\Cref{thm:upper-bound-adversarial}) }} \\
    \hline
    \textbf{\algNameNoSpace} & \multicolumn{2}{c|}{\makecell{$O(\sqrt{T \log T})$ \textbf{regret} \\ (\Cref{thm:full-info-stochastic,thm:partial-info-stochastic}) }} & \multicolumn{2}{c|}{\makecell{$(\rho / \vmax)$\textbf{-asymptotic} \\ \textbf{competitive ratio} \\ (\Cref{thm:full-info-adversarial,thm:partial-info-adversarial}) }} \\
    \hline
\end{tabular}

\bigskip

$T$: Number of auctions. \quad $\rho$: Average budget. \quad $\vmax$: Maximum value. 
    \caption{The bounds and the performance of \algName under different value input models and information structures.}
    \label{tab:summary-results}
\end{table}

In each round, our \algName algorithm is composed of three parts: (a) update the estimates on the target distribution according to previous samples; (b) decide the action according to the estimates and the pricing variable; (c) execute an online gradient descent (OGD) procedure on the pricing variable. 
Intuitively, the pricing variable measures whether the estimated cost is low compared to the estimated revenue given the value. 
The advertiser would choose to enter as long as the estimated reward is no less than the estimated cost after being scaled by the pricing parameter. 
After the decision, if the cost is low (or high, respectively) compared to the average budget, the pricing variable would correspondingly decrease (increase), rendering a higher (lower) probability for the advertiser to participate in the upcoming rounds. 
An advantage of such an adjustment method is that the total budget is used up smoothly around the target expenditure rate, which also fits the requirement of advertisers in practice. 
We should mention that a similar methodology has already been adopted in previous works, e.g., \cite{balseiro2022best}. 
Nevertheless, their approach does not deal with the underlying randomness of the highest competing bid given the value context. 

For the estimation part, we adopt the Dvoretzky–Kiefer–Wolfowitz (DKW) inequality \citep{massart1990tight} and form a confidence bound (CB) on the estimation results. 
With full information feedback, the number of samples the advertiser sees increases each round. As a result, the sum of the lengths of confidence intervals across all rounds can be bounded within $O(\sqrt{T\log T})$. 
Here, $T$ is the total number of auctions. 
Besides, with partial information feedback, it is important that the advertiser participates with a high frequency.
Or else the learning process will not converge quickly, and the algorithm's performance will have no guarantee. 
Via a novel argument on the dynamic change of the pricing variable (\Cref{lem:OGD-CB-entering-frequency}), we can show that our \algName algorithm guarantees a \emph{constant} entering frequency without any adjustment. 
Hence, the $O(\sqrt{T\log T})$ estimation bound still holds for our method with only partial information feedback. 

With the above skills, we show that under either full or partial information structure, our \algName algorithm reaches a near-optimal $O(\sqrt{T\log T})$ regret with stochastic $\bm v$ with probability $1 - O(1/T)$ (\Cref{thm:full-info-stochastic,thm:partial-info-stochastic}), i.e., an $O(\sqrt{T \log T})$ expected regret, and owns an optimal asymptotic competitive ratio with adversarial $\bm v$ (\Cref{thm:full-info-adversarial,thm:partial-info-adversarial}). 
We list these positive results together with the bounds and impossibility results in \Cref{tab:summary-results}. 

Apart from the high performance, our algorithm has two more advantages. First, our algorithm is computationally friendly. 
In each round, the update process of the pricing variable only takes a constant time. 
Second, compared with other algorithms, our solution does not rely on any particular properties of the distribution of the highest competing bid. (See \Cref{sec:related-work} for a literature review.) 
Specifically, our solution does not require the (interim) reward/cost function to be linear, convex/concave, or even continuous.
In particular, our solution even works for discrete distributions. 

\paragraph{Comparison between throttling and pacing.}
In \Cref{sec:comparison}, we compare the throttling strategy with the celebrated pacing strategy \citep{DBLP:journals/mansci/BalseiroG19} in repeated second-price auctions.
It is worth noting that the latter is known to be asymptotically optimal when $\bm v$ and $\bm p$ are simultaneously stochastic or adversarial.
When $\bm v$ and $\bm p$ are stochastic, we show that, in general cases, throttling results in a $\Theta(T)$ loss compared to pacing on the advertiser's expected revenue (\Cref{thm:linear-difference-throttling-pacing}). 
We also give special conditions under which these two strategies exhibit only an $\widetilde O(\sqrt{T})$ difference under full/partial information feedback (\Cref{thm:comparable-throttling-pacing}), for completeness. 
Excitingly, when $\bm v$ is adversarial and $\bm p$ is stochastic, we demonstrate that throttling is an \emph{asymptotically optimal} bidding strategy under full/partial information feedback.
Furthermore, our \algName algorithm is also optimal in this case. 
This result reveals the importance of throttling as a budget-smoothing method in advertising, and fills the gaps in research on dynamic bidding strategies with adversarial values and stochastic competing bids under repeated second-price auctions.

\subsection{Related Work} \label{sec:related-work}

In this part, we will review two popular budget management strategies in repeated auctions: throttling and pacing, further discussion on technically related works. 

Previous work on dynamic throttling mainly centers on experimental investigations.
Among them, \citet{DBLP:conf/wsdm/KarandeMS13} explore the concept of fair allocation in generalized second-price (GSP) auctions, wherein they present an optimal throttling algorithm for diverse objectives. 
\citet{DBLP:conf/kdd/AgarwalGWY14} also focus on GSP auctions from an advertiser's side and implements their algorithm in LinkedIn's ad serving system. 
\citet{DBLP:conf/kdd/XuLLQL15} evaluate a practical online throttling algorithm on the demand-side platform. 
Recently, \citet{gui2021auction} show how to conduct causal inference of online advertising effects in the budget throttling market. 
On the theoretical side, 
some work \cite{DBLP:journals/ior/BalseiroKMM21,DBLP:conf/wine/ChenKK21} focuses on the market equilibrium when all advertisers simultaneously follow a random throttling strategy from either a continuous or discrete view. 
In contrast, our work examines the dynamics of throttling in the repeated advertising market. 
Additionally, \citet{DBLP:conf/sigecom/CharlesCCDW13} study the regret-free allocation for advertisers' ROI, and shows that such a heuristic outperforms the random throttling strategy for advertisers. 
Meanwhile, other work looks into a similar problem for Internet keyword search, known as the AdWords problem \citep{DBLP:journals/jacm/MehtaSVV07,mehta2013online} in the framework of online matching. 
However, this line of work concentrates on the platform's side rather than the advertiser's side. 

Pacing is another well-studied budget control strategy, in which an advertiser shades her value by a constant factor on her bid. 
Existing work studies this strategy from both dynamic view \citep{DBLP:journals/mansci/BalseiroG19,DBLP:conf/www/BorgsCIJEM07,DBLP:journals/corr/abs-2205-08674,DBLP:conf/www/CelliCKS22,lee2013real} and equilibrium perspectives \citep{DBLP:journals/mansci/BalseiroBW15,DBLP:conf/ec/ConitzerKPSSMW19,DBLP:journals/ior/ConitzerKSM22}. 
Among these, the result of \citet{DBLP:journals/mansci/BalseiroG19} is highly correlated with our solution, which also considers the dual space. 
Nevertheless, the analysis of their algorithm depends on the continuity of the distribution function, which is not necessary for our algorithm. 
Some papers compare various budget control methods and explore their relationships from the equilibrium view \citep{DBLP:journals/ior/BalseiroKMM21,DBLP:journals/corr/abs-2203-16816,DBLP:journals/corr/abs-2102-10476}. 
In particular, \citet{DBLP:journals/ior/BalseiroKMM21} show that in the symmetric system equilibrium, throttling yields a higher profit for the platform than pacing under certain assumptions. 
A part of our results extends the comparison between throttling and pacing from an advertiser's viewpoint from a dynamic view.

Technically, our problem is closely related to the network revenue management (NRM) problem \citep{gallego1994optimal,gallego1997multiproduct} and the contextual bandits with knapsacks (CBwK) problem. 
However, there are significant differences between our work and the existing literature. 
In contrast to NRM, our agent's reward and cost are random with the highest competing bid, while such randomness is not present in NRM problems. 
Moreover, our setting diverges from the literature on CBwK in multiple ways.
Firstly, early work in CBwK \citep{DBLP:conf/colt/AgrawalDL16,DBLP:conf/colt/BadanidiyuruLS14} assumes that the context set is finite, whereas we do not make such assumptions. 
Secondly, some work in CBwK \citep{DBLP:conf/nips/AgrawalD16,sivakumar2022smoothed,han2022optimal,slivkins2022efficient} assumes a specific relationship between the expectation of reward/cost and the context, i.e., requires a specific distribution of the highest competing bid. 
However, our work sets no such limitation on this distribution. 
Finally, other work \citep{DBLP:conf/nips/WuSLJ15,balseiro2021survey,liu2022online,ai2022re} supposes that the context is drawn stochastically i.i.d. from some distribution. 
In contrast, our solution is oblivious to the context input model, allowing it to deal with adversarial context inputs. 
This is particularly important for auto-bidding services, as the value for an ad slot can be affected by multiple features and could vary over time. 
Therefore, it is crucial to design algorithms that can handle different value input models simultaneously. 
At last, our problem is also correlated with the bandits with knapsacks (BwK) problem (e.g., \cite{castiglioni2022online}). However, in that problem, the action in each round is chosen without any context, and the optimal action mode is universal. In contrast, in our problem, the optimal action is relevant to the value observed at the start of each auction. 

\section{Model}\label{sec:model}

\paragraph{Basic settings.} In this work, we consider the repeated second-price auction market, where an advertiser with a budget constraint competes against other advertisers. The market comprises $T$ rounds of auctions, and in each round $t \in [T] \coloneqq \{1, 2, \cdots, T\}$, an item is to be sold to a buyer via a second-price auction. Here, we suppose that $T \gg 1$. To match the more prevalent terminology in the literature, we refer to the ``advertiser'' as the ``buyer'' in the remaining parts. 

This work adopts the perspective of a fixed buyer. In each round $t \in [T]$, she obtains a personal value for the item, represented by $v_t \in [0, \vmax]$, where $\vmax$ is a constant upper bound on $v_t$.
We denote the highest competing bid against the buyer as $p_t$, which is assumed to be i.i.d. sampled from an unknown distribution $G$ with a support of $[0, \vmax]$. 
This assumption comes from the mean-field approximation \citep{DBLP:journals/mansci/BalseiroBW15} and is commonly used in the literature. 
We use $\bm v$ to represent the buyer's value vector across $T$ rounds, i.e., $\bm v \coloneqq (v_t)_{t \in [T]}$; similarly, $\bm p \coloneqq (p_t)_{t \in [T]}$. 

We assume the buyer has a total budget of $B$ across all $T$ rounds, with the maximum average expenditure being $\rho \coloneqq B/T$ per round. In this paper, we suppose that $\rho \leq \vmax$ is a constant. This assumption comes from the practice in which the buyer is always asked to set a budget for a fixed period, with relatively fixed rounds of auctions. 
In each round $t$, the buyer makes a decision $x_t \in \{0, 1\}$ based on the value $v_t$. The binary selection of the buyer reflects the nature of throttling, where $x_t = 1$ denotes participation in the auction, and $x_t = 0$ means saving the budget and not participating in the auction\footnote{It is required in throttling that the buyer truthfully bids. The motivation here is discussed in the Introduction.}. After the decision, the buyer receives a reward of $x_t r_t$ and incurs a cost of $x_t c_t$ in this round, where $r_t$ and $c_t$ are defined as: 
\begin{gather*}
    r_t = (v_t - p_t)^+, \quad c_t = p_t \mathbf 1[v_t \geq p_t]. 
\end{gather*}
In the above, $(v_t - p_t)^+$ in the expression for $r_t$ stands for the positive part of $(v_t - p_t)$. Thus, the buyer can obtain a positive reward and cost in round $t$ only by opting to ``enter'' (i.e., $x_t = 1$) and wins (i.e., $v_t \geq p_t$). In this case, her cost is $p_t$, and her reward is $v_t - p_t$ for a second-price auction. Once again, we mention here that in the literature on throttling, it is commonly assumed that the buyer bids truthfully as long as she enters an auction. 

\paragraph{Information structure.} 
Practically, the condition under which the buyer can observe $p_t$ of round $t$ is crucial to her strategy. Intuitively, the easier the buyer can view $p_t$, the more likely the buyer can obtain a throttling algorithm with a better total revenue guarantee. On this side, 
we consider two different information models in this work: 
\begin{enumerate}
    \item {\textbf{[Full information feedback.]}} The buyer observes $p_t$ at the end of any round $t$. 
    \item {\textbf{[Partial information feedback.]}} The buyer observes $p_t$ at the end of round $t$ only if she chooses to participate in the auction in this round, i.e., if $x_t = 1$. 
\end{enumerate}

In comparison to the full information feedback model, it is evident that the partial information feedback model is more challenging to manage since the buyer can access less information. 
Additionally, a natural model with even less information than the two we discuss is the bandit feedback model, in which the buyer only sees $r_t$ and $c_t$ in round $t$ instead of $p_t$. Recently, some research \citep{slivkins2022efficient,han2022optimal} has investigated this model in the problem of contextual bandits with knapsacks (CBwK), which is a generalization of our problem. Nevertheless, this line uses online regression techniques and has specific requirements on $\mathbb{E}[r_t, c_t \mid v_t]$ as a function of $v_t$, e.g., being linear. In other words, their method has strong assumptions on the distribution $G$ of $p_t$ in our problem. As far as we know, these assumptions are \emph{necessary} in the literature for bandits information feedback. In contrast, in this work, we do not impose any restriction on the distribution $G$. Our solution is \emph{oblivious} to the distribution $G$, and thus model-free. In this respect, our information feedback model is comparable to existing works. 

We generally use $\mathcal H_t$ to denote the history that the buyer can access at the start of round $t$. For the full information feedback model, we use $\mathcal H_t^{\rm F} \coloneqq (v_\tau, x_\tau, p_\tau)_{1\leq \tau < t}$ to denote the buyer's view at the start of round $t$. 
We should note that in this model, $p_\tau$ is always available to the buyer at the end of each round $\tau$, and $r_\tau$ and $c_\tau$ can be inferred from $v_\tau$, $x_\tau$ and $p_\tau$. For the partial information feedback model, since $p_\tau$ is disclosed to the buyer if and only if $x_\tau = 1$, we accordingly define the history available at round $t$ as $\mathcal H_t^{\rm P}\coloneqq (v_\tau, x_\tau, x_\tau p_\tau)_{1\leq \tau < t}$. 
It is worth noting that $p_\tau$ cannot be deduced from $x_\tau$ and $x_\tau p_\tau$ when $x_\tau = 0$. Likewise, $r_\tau$ and $c_\tau$ can also be derived using $v_\tau$, $x_\tau$ and $x_\tau p_\tau$ in this information model. 

\paragraph{The throttling strategy.} With the above notation, we now formally define the buyer's throttling strategy. Prior to making a decision, the buyer can see $\mathcal H_t$ and $v_t$ in each round $t$. Denoting the buyer's single-round strategy in round $t$ by $\beta_t: [0, \vmax]^{2t - 1} \times \{0, 1\}^{t - 1} \times \Gamma \to \{0, 1\}$, we have
\[x_t = \beta_t\left(\mathcal H_t, v_t, \gamma_t\right). \]
Here, $\gamma_t\in \Gamma$ is sampled from a probability space to depict the potential randomness involved in calculating $x_t$. Let $\bm \gamma = (\gamma_t)_{t \in [T]}$.
As a result, $\beta \coloneqq (\beta_t)_{t \in [T]}$ encompasses the buyer's single-round strategy for all $T$ rounds and represents her overall throttling strategy. 

With $\bm v$, $\bm p$ and randomness $\bm \gamma$ given, we denote the stopping time of strategy $\beta$ by $T_0^{\beta}(\bm v, \bm p, \bm \gamma) \leq T$, i.e., the last round with $x_t = 1$. When the context is clear, we abbreviate this as $T_0$. Consequently, the total revenue of $\beta$ with inputs $\bm v$ and $\bm p$ is given by: 
\[\totalRevenueGamma \coloneqq \sum_{t = 1}^{T_0} x_t r_t. \]

\subsection{Value Model and Benchmark}

As previously suggested, this work examines two distinct input models for the values $(v_t)_{t \in [T]}$: the stochastic model and the adversarial model.

\paragraph{Stochastic values and regret.} 
Concerning the stochastic value model, it is assumed that for each $t$, $v_t$ is drawn i.i.d. from some \emph{unknown} distribution $F$ with a support on $[0, \vmax]$. We additionally suppose that $F$ and $G$ are independent. 
In this case, the performance of a throttling strategy is assessed by comparing its reward with the fluid adaptive throttling benchmark. The latter represents the optimal expected revenue of any random strategy given the value without exceeding the budget in expectation. Specifically, 
\begin{gather}\label{eq:OPT}
    \begin{gathered}
    \OPT \coloneqq T\cdot \max_{\pi: [0, \vmax] \to [0, 1]} \mathbb E_{v \sim F, p \sim G} \left[\pi(v) \cdot (v - p)^+\right], \\
    \quad {\rm s.t.}\quad \mathbb E_{v \sim F, p \sim G} \left[\pi(v) \cdot p \mathbf 1[v\geq p]\right] \leq \rho.
    \end{gathered}
\end{gather}
It is known that $\OPT$ provides an upper bound for the expected total reward of any throttling strategy with stochastic values \citep{balseiro2021survey}. 
Consequently, we can define the regret of strategy $\beta$ in relation to $\OPT$ given $\bm v$, $\bm p$ and randomness $\bm \gamma$. That is, 
\[\regretGamma \coloneqq \OPT - \totalRevenueGamma. \]
Clearly, under stochastic values, our goal is to design a throttling strategy $\beta$ that results in a low expectation of $\regretGamma$. A stronger requirement is to ensure a small $\regretGamma$ with high probability on samples $(\bm v, \bm p)$ and randomness $\bm \gamma$. 

\paragraph{Adversarial values and asymptotic competitive ratio.} 
We also consider the scenario that the inputs of $(v_t)_{t \in [T]}$ are adversarial on $[0, \vmax]$, which accounts for the situations where the buyer lacks confidence in the item distribution. In this regard, we first define the hindsight throttling benchmark, which is the optimal performance advertiser could attain with the benefit of hindsight on $\bm v$ and $\bm p$. 
Specifically, 
\begin{gather}\label{eq:OPT-hindsight}
    \begin{gathered}
    \totalRevenueHindsight\coloneqq\max _{\bm x \in\{0,1\}^T} \sum_{t=1}^T x_{t}\left(v_{t}-p_{t}\right)^+, \\ \quad {\rm s.t.}\quad  \sum_{t=1}^T x_{t} p_{t} \mathbf 1[v_t\geq p_t] \leq T \rho.
    \end{gathered}
\end{gather}

Evidently, $\totalRevenueHindsight$ bounds $\totalRevenueGamma$ for any strategy $\beta$ and randomness $\bm \gamma$. We then define a throttling strategy $\beta$ to be asymptotically $\mu$-competitive for $\mu \in (0, 1]$, if the following condition holds: 
{
\small
\[
   \liminf_{T\rightarrow \infty, B = \rho T} \inf_{\bm v, G} \left(\frac{1}{T}\cdot \mathbb{E}_{\bm p\sim G^T, \bm \gamma}\left[\totalRevenueGamma - \mu\cdot  \totalRevenueHindsight\right]\right)\geq 0. 
\]
}




\section{Bounds and Impossibility Results}\label{sec:b-i-r}

In this section, we study the best possible performance of any throttling strategy for either stochastic or adversarial values. In the case of stochastic values, we obtain a lower bound of $\Omega(\sqrt{T})$ regret. On the other hand, with adversarial values, we give an upper bound asymptotic competitive ratio of $\rho / \vmax$ for any throttling algorithm. We further derive an impossibility result indicating that any throttling algorithm could achieve arbitrarily small reward facing adversarial $\bm p$ even when $\{v_t\}_{t \in [T]}$ is fixed. This result states the necessity of supposing stationary highest competing bids in the throttling context. Readers can refer to the omitted proofs in this section in \Cref{sec:proof-b-i-r}. 

\subsection{Stochastic Values}

Our main result for the regret lower bound with stochastic values is the following theorem: 

\begin{theorem}\label{thm:lower-bound-stochastic}
    There exists an instance tuple $(F, G)$ and some constant $\constantLowerBound > 0$, such that for any online throttling strategy $\beta$ and $4 | T$ (i.e., $T$ is a multiple of $4$), we have
    \[\mathbb E_{\bm v\sim F^T, \bm p \sim G^T, \bm \gamma} \left[\regretGamma\right] \geq \constantLowerBound\sqrt{T}. \]
\end{theorem}

Similar results have already existed in the literature of revenue management \citep{DBLP:journals/mansci/VeraB21,DBLP:journals/mansci/BumpensantiW20,arlotto2019uniformly}. Nevertheless, in the revenue management problem, there is no randomness in the reward and cost given the context, which is a simplification of our situation. Here, we provide a proof exclusive to our setting. The central idea lying in the proof is to give a proper problem instance with a \emph{degenerate} fluid solution, under which even the optimal throttling strategy with the hindsight of $\bm v$ and $\bm p$ does not escape a regret of $\Omega(T)$. 

Specifically, we suppose $\vmax = 1$ and $\rho = 1/2$, and the instances for other $(\vmax, \rho)$ pairs can be correspondingly constructed. In this case, we consider the problem instance with $v$ always equal to 1, and $G$ be the uniform binary distribution on $1/3$ and $2/3$. Therefore, the resulting programming $\OPT$ is a degenerate LP. We start by establishing an upper bound on $\totalRevenueGeneral$ for any online throttling strategy $\beta$ when $4|T$. Then we give a combinatorial formula for the lower bound on regret in terms of $\OPT$. Finally, we simplify this lower bound and use Stirling's formula to achieve the result. 

Note that the performance of an online throttling strategy could only be worse with partial information. Therefore, \Cref{thm:lower-bound-stochastic} also implies that no online throttling strategy can obtain a better regret in hindsight with partial information feedback. Together with \Cref{thm:full-info-stochastic} and \Cref{thm:partial-info-stochastic}, this establishes the near-optimality of our \algoName algorithm in both information settings.

\subsection{Adversarial Values}

In this part, we consider the model when $\bm v$ is adversarial. In this case, we settle an upper bound on the asymptotic competitive ratio of any throttling algorithm. In fact, we have the following theorem stating that such upper bound is $\rho / \vmax$. 
\begin{theorem}\label{thm:upper-bound-adversarial}
    For any $\mu > \rho / \vmax$, 
    \[
       \liminf_{T\rightarrow \infty, B = \rho T} \inf_{\bm v, G}\left(\frac{1}{T}\cdot \mathbb{E}_{\bm p\sim G^T, \bm \gamma}\left[\totalRevenueGamma - \mu\cdot  \totalRevenueHindsight\right]\right) < 0. 
    \]
\end{theorem}

Here we provide intuition on the proof, which follows \cite{DBLP:journals/mansci/BalseiroG19}. More specifically, we first apply Yao's principle \citep{yao1977probabilistic} and change our problem into constructing a ``hard'' problem instance with stochastic $\bm p$ and half-stochastic $\bm v$ that follows some vector distribution. Here, the ``hard'' means that the asymptotic competitive ratio of any \emph{deterministic} (rather than random) algorithm is no more than $\rho / \vmax$ under the instance. We further let $\bm p$ be fixed across time. Therefore we are further reduced to constructing a vector distribution of $(\bm v - \bm p)^+$ that blocks any deterministic throttling strategy. 

For such a vector distribution, we aim to ``cheat'' the buyer to fall into unsatisfying chances. On this side, supports of the distribution are utility vectors that have an identical and constant-fraction length low-revenue prefix. For the suffix, the utilities vary largely across the supports. Such a construction results in a dilemma for any throttling strategy or even bidding strategy: If the strategy chooses to win much in the prefix, then it suffers from the underlying low revenue. On the other hand, if the strategy chooses to bet on the suffix and give up the prefix, then the accumulated revenue has the probability to be either very high or very low (even a lot worse than the prefix) and is largely uncertain. Our construction follows this idea and further adopts a fine-tuning of parameters to achieve the upper bound result.

\subsection{An Impossibility Result}\label{sec:impossibility-result}

An important question that could be raised is the behavior of a throttling strategy when the highest competing bid $\bm p$ is adversarial. As a completion of the previous lower/upper bound results, the following result shows that no algorithm can guarantee asymptotic $\mu$-competitiveness for any $\mu>0$ under adversarial $\bm p$, even when $v_t$ is fixed for each round.
\begin{theorem}\label{thm:impossibility-result}
   For any online throttling strategy $\beta$, randomness $\bm \gamma$, $\mu>0$ and $T$, there exists an instance tuple $(\bm v,\bm p)$ for each $T > 0$, such that 
    \[\frac{1}{T}\left(\totalRevenueGamma - \mu \cdot \totalRevenueHindsight\right) < 0.\]
\end{theorem}

Intuitively, the result follows since the buyer has no idea about $p_t$ giving $v_t$. Further, the buyer's choice is limited to binary actions. Therefore, it is possible that the buyer takes a high cost to win an insufficient auction while losing cheap chances. Therefore, the case that $\bm p$ is adversarial is impossible for the throttling strategy. 
With the theorem, it is natural that we suppose that $\{p_t\}_{t \in [T]}$ follows an i.i.d. distribution on the theoretical side.

\section{The \algName Algorithm and Performance}\label{sec:algorithm-performance}

\begin{algorithm}[!ht]
    \SetKwInput{KwInit}{Initialization}
    \SetKw{break}{break}
    \KwIn{$\rho, T$.}
    \KwInit{$\mathcal I_1 \gets \emptyset$, $B_1 \gets B$, $\lambda_1 \gets 0$.}
    \BlankLine
    \For{$t \gets 1$ \KwTo $T$}{
        Observe $v_t$\;
        \BlankLine
        \tcc{A single round of exploration.}
        \If{$t = 1$}{\label{algoline:explore_start}
            $x_t \gets 1$\;
            $\lambda_{t+1} \gets \lambda_t$\;
        }\label{algoline:explore_end}
        \BlankLine
        \Else{
            \BlankLine
            \tcc{Estimate the revenue and cost with a confidence bound.}
            $\epsilon_t \gets \sqrt{(\ln 2 + 2\ln T) / (2|\mathcal I_t|)}$\; 
            $\estr_t(v_t) \gets (\sum_{\tau \in \mathcal I_t} (v_{t} - p_\tau)^+) / |\mathcal I_t| + \epsilon_tv_t$,
            $\estc_t(v_t) \gets (\sum_{\tau \in \mathcal I_t} p_\tau \mathbf 1[v_{t} \geq p_\tau]) / |\mathcal I_t|- 2\epsilon_tv_t$\; \label{algoline:estimate}
            \BlankLine
            \tcc{Choose the action according to the estimates.}
            $x_t \gets \mathbf 1[\estr_t(v_t) \geq \lambda_t\estc_t(v_t)]$\; \label{algoline:action}
            \BlankLine
            \tcc{Online gradient descent on the pricing variable.}
            $\eta_t \gets 1 / (\vmax \sqrt{t})$\;
            $\lambda_{t + 1} \gets \left(\lambda_t + \eta_t (x_t\estc_t(v_t) - \rho)\right)^+$\; \label{algoline:grad_descent}
        }
        \BlankLine
        \tcc{Observe the sample.}
        \If{(\textit{FULL-INFO}) $\vee$ (\textit{PARTIAL-INFO} $\wedge\, x_t = 1$) \label{algoline:observe_condition}}
        {
            Observe $p_t$\;
            $\mathcal I_{t+1} \gets \mathcal I_t \cup \{t\}$\;
        }
        \Else{
            $\mathcal I_{t+1} \gets \mathcal I_t$\;
        }
        \BlankLine
        \tcc{Update the remaining budget.}
        $B_{t + 1} \gets B_t - x_t c_t$\;
        \If{$B_{t+1} < \vmax$}
        {
            \break\;
        }
    }
    \caption{\algName}
    \label{alg:OGD-CB}
\end{algorithm}

In this section, we introduce an online throttling strategy known as \algNameNoSpace, which is oblivious of the value model (stochastic or adversarial value) and the information structure (full or partial information feedback). 
For stochastic values, we obtain an upper bound of $O(\sqrt{T\log T})$ on the regret of the \algName algorithm with either full information or partial information feedback, which is near optimal considering \Cref{thm:lower-bound-stochastic}.
For adversarial values, we show that the \algName algorithm is asymptotically $(\rho / \vmax)$-competitive regardless of the information model, which matches the lower bound given in \Cref{thm:upper-bound-adversarial}. 
Specifically, we start with an introduction to the algorithm (\Cref{sec:algorithm}). We consider the full information setting for stochastic and adversarial values, respectively, in \Cref{sec:performance-stochastic} and \Cref{sec:performance-adversarial}. We further give the results under the partial information setting in \Cref{sec:performance-partial-info}. 
All omitted proofs in this section can be found in \Cref{sec:proof-algo}. 

\subsection{The Algorithm}\label{sec:algorithm}

Our \algName algorithm is presented in \Cref{alg:OGD-CB}. The algorithm starts with a one-round exploration to make an appropriate initialization (\Cref{algoline:explore_start}--\Cref{algoline:explore_end}). In each of the following rounds, after observing the value, the algorithm chooses the action based on a dynamically updated pricing parameter $\lambda_t$ (\Cref{algoline:action}), which is updated to control the rate of budget expenditure (\Cref{algoline:grad_descent}). The update of $\lambda_t$ follows an online gradient descent (OGD) procedure for a series of proper online reward functions, with step size $\eta_t$. Intuitively, a large $\lambda_t$ indicates that the budget is being spent too quickly, and the algorithm reduces the frequency of entering the market. Conversely, an average expenditure below the ideal $\rho$ in past rounds will result in a descent of $\lambda_t$ and encourage the algorithm to participate in the auction. 
This intuition has been inspired by \citet{balseiro2022best}. However, a crucial distinction between our setting and that work is that the buyer is unaware of the (expected) revenue and cost given the value. To address this issue, we employ the distribution estimation method and the confidence bound (CB) technique. Specifically, at the start of each round, the algorithm first provides estimates of the expected revenue and cost based on the history (\Cref{algoline:estimate}) and incorporates a bias using the confidence bound, parameterized by $\epsilon_t$, and then makes the decision and updates according to the biased estimation. As the observation on the highest competing bid $p$ accumulates, the estimation of the reward and the cost becomes more precise, and the bias value reduces to zero.

\subsection{Stochastic Values with Full Information Feedback}\label{sec:performance-stochastic}

We now analyze the performance of the \algName algorithm with stochastic values and full information feedback. We obtain that in this scenario, \algName achieves an $O(\sqrt{T\log T})$ regret bound against $\opt$, as given in the following theorem.
\begin{theorem}\label{thm:full-info-stochastic}
    With full information feedback, when $\{v_t\}_{t \in T}$ is sampled i.i.d. from some distribution $F$ on $[0, \vmax]$, there is a constant $\constantFull$, such that with probability at least $1 - 4/T$, the \algName algorithm guarantees
    \begin{gather*}
        \regret \leq \constantFull\sqrt{T \log T}. 
    \end{gather*}
\end{theorem}

To prove the theorem, we first need some concentration bounds for the estimation process in \algNameNoSpace. Below we show that $\estr_t(v_t)$ and $\estc_t(v_t)$ are good estimations of $r(v_t) \coloneqq \mathbb E_{p\sim G}\left[(v_t - p)^+\right]$ and $c(v_t) \coloneqq \mathbb E_{p\sim G}\left[p\mathbf 1[v_t \geq p]\right]$ correspondingly, by applying Dvoretzky-Kiefer-Wolfowitz (DKW) inequality \citep{massart1990tight}. 

\begin{lemma}\label{lem:estimation-r-c}
    With full information feedback, in round $1 < t \leq T_0$, with failure probability $q(t, \epsilon_t) \coloneqq 2\exp (-2(t - 1)\epsilon_t^2)$, for any $v_t \in [0, \vmax]$, we have
    \begin{gather}
        r(v_t)\leq \estr_t(v_t)\leq r(v_t) + 2\epsilon_tv_t, \label{eq:r-estimation}\\
        c(v_t) - 4\epsilon_tv_t\leq \estc_t(v_t)\leq c(v_t). \label{eq:c-estimation} 
    \end{gather}
\end{lemma}

Another crucial ingredient of the proof concerns with the optimization process in \Cref{algoline:grad_descent} of \Cref{alg:OGD-CB}. \Cref{lem:lambda_upper_bound} gives a useful observation about $\lambda_t$.

\begin{lemma}\label{lem:lambda_upper_bound}
    For any $t \leq T_0 + 1$, we have $\lambda_t \in [0, \vmax / \rho - 1]$.
\end{lemma}

With \Cref{lem:lambda_upper_bound} in hand, a standard analysis of the online gradient descent method \citep{hazan2016introduction} with online gain function $h_t(\lambda) = \lambda  (x_t\estc_t(v_t) - \rho)$ gives the following result:

\begin{lemma}\label{lem:OGD}
    For any $\lambda \in [0, \vmax / \rho - 1]$, we have
    \begin{equation}\label{eqn:OGD}
        \sum_{t=2}^{T_0} \lambda_t (x_t \estc_t(v_t) - \rho) \geq \lambda \sum_{t=2}^{T_0} (x_t \estc_t(v_t) - \rho) - \left(\frac{\left(\vmax / \rho - 1\right)^2}{\eta_{T_0}} + \vmax^2 \sum_{t=2}^{T_0} \eta_t\right).
    \end{equation}
\end{lemma}

With the preparations, we provide a high-level proof sketch of \Cref{thm:full-info-stochastic}. 
\begin{proof}[Proof Sketch of \Cref{thm:full-info-stochastic}]
We prove the result in four steps. 
\paragraph{Step 1.} We first lower bound the performance of our throttling strategy by using Azuma--Hoeffding inequality and \Cref{lem:estimation-r-c}. We arrive at
\[
    \totalRevenueGeneral \geq \sum_{t=2}^{T_0}x_t\estr_t(v_{t}) - O\left(\sqrt{T \log T}\right).
\]

\paragraph{Step 2.} We next bound $\sum_{t=2}^{T_0}x_t\estr_t(v_{t})$ by rewriting it as 
\[
    \sum_{t=2}^{T_0}x_t\estr_t(v_{t}) = \underbrace{\sum_{t=2}^{T_0} \left(x_t\estr_t(v_t) - \lambda_t \left(x_t\estc_t(v_t) - \rho\right)\right)}_{R_1} + \underbrace{\sum_{t=2}^{T_0} \lambda_t\left(x_t\estc_t(v_t) - \rho\right)}_{R_2}, 
\]
and bounding $R_1, R_2$ respectively. It is worth mentioning that $R_1$ is similar to the objective of the dual problem for programming $\OPT$, which intuitively implies $R_1$ is ``close'' to $(T_0 - 1)\OPT / T$. By applying Azuma--Hoeffding inequality and \Cref{lem:estimation-r-c} again, we obtain 
\[
    R_1 \geq \frac{T_0 - 1}{T} \OPT - O\left(\sqrt{T \log T}\right).
\]
As for $R_2$, we apply \Cref{lem:OGD} to obtain 
\[
    R_2 \geq - O\left(\sqrt{T}\right).
\]

\paragraph{Step 3.} The third step proves that the ending time $T_0$ of our strategy is close to $T$. More precisely, we show that
\[
    \frac{T_0 - 1}{T}\OPT \geq \OPT - O\left(\sqrt{T \log T}\right).
\]

\paragraph{Step 4.} Finally, the proof is completed by putting everything together and using a union bound to lower bound the total failure probability. 
\end{proof}

Clearly, by taking an expectation, a direct corollary of \Cref{thm:full-info-stochastic} is that the expected regret of the \algName algorithm is $O(\sqrt{T \log T})$. 
\begin{corollary}\label{coro:full-info-stochastic}
    With full information feedback, when $\{v_t\}_{t \in T}$ is sampled i.i.d. from some distribution $F$ on $[0, \vmax]$, the \algName algorithm guarantees
    \begin{gather*}
        \mathbb E_{\bm v, \bm p}\left[\regret\right] = O\left(\sqrt{T \log T}\right). 
    \end{gather*}
\end{corollary}

\subsection{Adversarial Values with Full Information Feedback}\label{sec:performance-adversarial}

In this section, we consider the case that the value input could be adversarial and the highest competing bid is stochastic. In this case, we show that the \algName algorithm has a $\rho$ asymptotic competitive ratio, which matches the upper bound given in \Cref{thm:upper-bound-adversarial}. 

\begin{theorem}\label{thm:full-info-adversarial}
    With full information feedback, the \algName algorithm guarantees
    \[
    \liminf_{T\rightarrow \infty, B = \rho T} \inf_{\bm v, G} \left(\frac{1}{T}\cdot \mathbb{E}_{\bm p\sim G^T}\left[\totalRevenueGeneral - \frac{\rho}{\vmax} \cdot \totalRevenueHindsight\right]\right)\geq 0. 
    \]
\end{theorem}

\begin{proof}[Proof Sketch of \Cref{thm:full-info-adversarial}]
We prove the result in five steps.

\paragraph{Step 1.}
We first give a rough bound of the performance of our throttling strategy with estimation errors and the OCO regret as follows:
\begin{equation*}
    \mathbb{E}_{\bm p\sim G^T}\left[\sum_{t=2}^{T_0} x_t\estr_t(v_t)\right]\geq \frac{\rho}{\vmax}\cdot \mathbb{E}_{\bm p\sim G^T} \left[\sum_{t=2}^{T_0}\estr_t(v_t)\right]+ \mathbb{E}_{\bm p\sim G^T}\left[\sum_{t=2}^{T_0}\lambda_t(x_t\estc_t(v_t)-\rho)\right].
\end{equation*}

\paragraph{Step 2.}
We next bound the OCO regret. Using \Cref{lem:OGD}, we derive
\begin{equation*}
    \mathbb{E}_{\bm p\sim G^T}\left[\sum_{t=2}^{T_0}\lambda_t(x_t\estc_t(v_t)-\rho)\right]\geq -O\left(\sqrt{T}\right).
\end{equation*}

\paragraph{Step 3.}
The third step bounds the difference between $\sum_{t=2}^{T_0}x_t\estr_t(v_t)$ and $\totalRevenueGeneral$, $\sum_{t=2}^{T_0}\estr_t(v_t)$ and $\totalRevenueHindsight$ while keeping the difference between $T$ and $T_0$. More precisely, we use \Cref{lem:estimation-r-c} and show that
\begin{equation*}
    \mathbb{E}_{\bm p\sim G^T}\left[\sum_{t=2}^{T_0}x_t\estr_t(v_t)\right]\leq \mathbb{E}_{\bm p\sim G^T}\left[{\totalRevenueGeneral}\right]+O\left(\sqrt{T\log T}\right)
\end{equation*}
and 
\begin{equation*} 
    \mathbb{E}_{\bm p\sim G^T}\left[\sum_{t=2}^{T_0}\estr_t(v_t)\right]\geq \mathbb{E}_{\bm p\sim G^T}\left[\totalRevenueHindsight\right]-(T-T_0+1)\cdot \vmax-O\left(1\right). 
\end{equation*}

\paragraph{Step 4.}
In this step, we further bound the difference between $T_0$ and $T$ and obtain that
\begin{equation*}
     T-T_0+1\leq O\left(\sqrt{T\log T}\right).
\end{equation*}

\paragraph{Step 5.}
Finally, the proof is completed by putting everything together.
\end{proof}

\subsection{Partial Information Feedback}\label{sec:performance-partial-info}

In the partial information setting, the buyer can observe $p_t$ only if she enters the round $t$. we first give an analog of \Cref{lem:estimation-r-c} in the partial information feedback setting without proof. 
\begin{lemma}\label{lem:estimation-r-c-partial}
    With partial information feedback, in round $t \geq 1$ with failure probability $q'(t, \epsilon_t) := 2\exp (-2|\mathcal I_t|\epsilon_t^2)$, for any $v_t \in [0, \vmax]$, we have
    \begin{gather*}
        r(v_t)\leq \estr_t(v_t)\leq r(v_t) + 2\epsilon_tv_t, \\
        c(v_t) - 4\epsilon_tv_t\leq \estc_t(v_t)\leq c(v_t). 
    \end{gather*}
\end{lemma}

The main challenge of partial information feedback is that the algorithm may not collect enough historical data to give good estimates of $r(v_t)$ and $c(v_t)$. In other words, we must simultaneously limit the failure probability and the estimation error.
We overcome this challenge by bounding the entering frequency. 
Specifically, we use an induction method to show that $|\mathcal I_t|$, the entering frequency before round $t$, grows with $t$ linearly. The result is formalized in the following important lemma. 

\begin{lemma}\label{lem:OGD-CB-entering-frequency}
	Let $\constantFrequency := \min\{(1/2)\cdot (\rho / \vmax)^2, (\sqrt{2}/4)\cdot (\rho / \vmax)\}$. In the partial information setting, for any $t \geq 2$, the \algName algorithm guarantees that $|\mathcal I_t| \geq \constantFrequency\cdot (t - 1)$.
\end{lemma}

With the help of \Cref{lem:estimation-r-c-partial} and \Cref{lem:OGD-CB-entering-frequency}, we can bound the accumulated inaccuracy in estimating $r(v_t)$ and $c(v_t)$ with partial information feedback. Therefore, we can revisit \Cref{thm:full-info-stochastic,thm:full-info-adversarial} and derive their counterparts under partial information. Specifically, we have the following results. 

\begin{theorem}\label{thm:partial-info-stochastic}
    With partial information feedback, when $\{v_t\}_{t \in T}$ is sampled i.i.d. from some distribution $F$ on $[0, \vmax]$, there is a constant $\constantPartial$, such that with probability at least $1 - 4/T$, the \algName algorithm guarantees
    \begin{gather*}
        \regret \leq \constantPartial\sqrt{T \log T}. 
    \end{gather*}
\end{theorem}

\begin{corollary}\label{coro:partial-info-stochastic}
    With partial information feedback, when $\{v_t\}_{t \in T}$ is sampled i.i.d. from some distribution $F$ on $[0, \vmax]$, the \algName algorithm guarantees
    \begin{gather*}
        \mathbb E_{\bm v, \bm p}\left[\regret\right] = O\left(\sqrt{T \log T}\right). 
    \end{gather*}
\end{corollary}

\begin{theorem}\label{thm:partial-info-adversarial}
    With partial information feedback, the \algName algorithm guarantees
    \[
    \liminf_{T\rightarrow \infty, B = \rho T} \inf_{\bm v, G} \left(\frac{1}{T}\cdot \mathbb{E}_{\bm p\sim G^T}\left[\totalRevenueGeneral - \frac{\rho}{\vmax} \cdot \totalRevenueHindsight\right]\right)\geq 0. 
    \]
\end{theorem}

We notice that in \Cref{thm:full-info-stochastic,thm:partial-info-stochastic}, the difference between $\constantPartial$ and $\constantFull$ satisifies that
\[\constantPartial - \constantFull = O\left(\vmax\cdot \sqrt{\max\left\{\frac{2\vmax^2}{\rho^2}, \frac{4\vmax}{\sqrt{2}\rho}\right\}}\right).\]
In other words, the difficulty of partial information feedback highly correlates with the ratio $\vmax / \rho$. 
When this ratio increases, i.e., the buyer's average budget $\rho$ becomes smaller, the partial information feedback becomes more challenging. 
Intuitively, with smaller $\rho$, the buyer has fewer chances to participate in auctions and thus learns less information on the distribution of the highest competing bid $p$ with partial feedback, which leads to an increase in regret.

\section{Comparison between Throttling and Pacing} \label{sec:comparison}

In this section, we focus on comparing two popular budget control methods in the dynamic setting. The first one is throttling, which is studied in this work. The second is pacing \citep{DBLP:journals/mansci/BalseiroG19}, in which the buyer can shade her value by an adaptive multiplier and bid the shaded value in each round. This approach is extensively studied in literature and widely adopted in the industry, as it is shown to be the asymptotically optimal bidding strategy when $\bm v$ and $\bm p$ are both stochastic or adversarial. Specifically, we are most interested in comparing the two dynamic strategies. This part extends the result in \cite{DBLP:journals/ior/BalseiroKMM21}, which compares these two strategies in system equilibrium, i.e., when the dynamic process converges. 

Before the mathematical analysis, we first clarify that the pacing strategy is supported by a continuum action space, as the buyer is free to bid any value under the model. In contrast, the action space for throttling is binary. Further, blessed by its solution structure, the adaptive pacing strategy works with bandits information; that is, the algorithm only receives the reward and cost each round rather than the highest competing bid. This advantage is due to the matching between the pacing strategy and the dual problem of hindsight programming. Therefore, the pacing approach does not involve any distribution learning technique. As a comparison, such an effect is not achievable for the throttling strategy due to the limited action space, or on a larger scale, the problem of contextual bandits with knapsacks without a fine structure among the context, revenue, and cost. 

We now compare the two strategies under stochastic and adversarial values. Specifically, we restrict to full or partial information feedback model. We use $\totalRevenueThrottling$ to represent the revenue of the optimal throttling strategy\footnote{The ``optimality'' here concerns the expected total revenue. When $\bm v$ is stochastic, the expectation is taken on $\bm v$, $\bm p$ and the algorithm randomness $\bm \gamma$. When $\bm v$ is adversarial, the expectation is only on $\bm p$ and $\bm \gamma$.} given $\bm v$, $\bm p$, and $\bm \gamma$. As a comparison, we use $\totalRevenuePacing$ to represent the revenue of the adaptive pacing strategy given in \cite{DBLP:journals/mansci/BalseiroG19}. 

\subsection{Stochastic Values}

We first consider the case with stochastic values. In general, it is hard to compare the performance of two dynamic strategies. Therefore, a natural idea is to resort to two static benchmarks as intermediates that give good approximations of these two dynamics, respectively, and then compare these two static values. We show that such an approach is realizable. We start by recalling the fluid benchmark as given by \eqref{eq:OPT}: 
\begin{gather*}
    \OPT \coloneqq T\cdot \max_{\pi: [0, \vmax] \to [0, 1]} \mathbb E_{v \sim F, p \sim G} \left[\pi(v) \cdot (v - p)^+\right], \quad {\rm s.t.}\quad \mathbb E_{v \sim F, p \sim G} \left[\pi(v) \cdot p \mathbf 1[v\geq p]\right] \leq \rho. 
\end{gather*}
As given by \Cref{thm:full-info-stochastic,thm:partial-info-stochastic}, we have the following proposition bounding on the revenue of the optimal throttling strategy with full or partial information feedback: 
\begin{proposition}\label{prop:throttling-stochastic-lower-upper}
    For any problem instance $(F, G)$, under full or partial information feedback, we have: 
    \[\OPT - O\left(\sqrt{T \log T}\right) \leq \mathbb E_{\bm v, \bm p, \bm \gamma}\left[\totalRevenueThrottling\right] \leq \OPT. \]
\end{proposition}

In other words, $\OPT$ gives a near-accurate approximation of the optimal throttling strategy. As a comparison, the adaptive pacing strategy given by \cite{DBLP:journals/mansci/BalseiroG19} reaches the following performance, which is near optimal: 
\begin{proposition}[From \cite{DBLP:journals/mansci/BalseiroG19}]\label{prop:pacing-stochastic-upper}
    For any problem instance $(F, G)$, we have
    \[\mathbb E_{\bm v, \bm p}\left[\totalRevenueHindsight - \totalRevenuePacing\right] = O\left(\sqrt{T}\right). \]
\end{proposition}
Here, as presented in \eqref{eq:OPT-hindsight}, the hindsight benchmark is given by the following: 
\begin{gather*}
    \totalRevenueHindsight\coloneqq\max _{\bm x \in\{0,1\}^T} \sum_{t=1}^T x_{t}\left(v_{t}-p_{t}\right)^+, \quad {\rm s.t.}\quad  \sum_{t=1}^T x_{t} p_{t} \mathbf 1[v_t\geq p_t] \leq T \rho.
\end{gather*}
Therefore, to compare the performance of these two widely adopted dynamic strategies, it suffices for us to reason the difference between the two benchmarks $\OPT$ and $\mathbb E_{\bm v, \bm p}[\totalRevenueHindsight]$ when disregarding the constants on the $\widetilde O(\sqrt{T})$ term. 
Nevertheless, note that $\totalRevenueHindsight$ is the hindsight benchmark and is hard to be directly put together with the fluid benchmark $\OPT$. On this side, we further introduce a bidirectional fluid benchmark $\OPTstrong$, also known as deterministic LP in the literature, defined as follows: 
\begin{gather*}
    \OPTstrong \coloneqq T\cdot \max_{\kappa: [0, \vmax]^2 \to [0, 1]} \mathbb E_{v \sim F, p\sim G} \left[\kappa(v, p)\cdot (v - p)^+\right], \quad {\rm s.t.}\quad \mathbb E_{v \sim F, p\sim G} \left[\kappa(v, p)\cdot p \mathbf 1[v \geq p] \right] \leq \rho. 
\end{gather*}

As revealed by fruitful literature on the problem of revenue management, $\OPTstrong$ sets a satisfying approximation of the hindsight benchmark from the upper side.
\begin{proposition}[From \cite{gallego1994optimal,gallego1997multiproduct,talluri1998analysis,DBLP:journals/mansci/VeraB21,DBLP:journals/mansci/BumpensantiW20}]\label{prop:hindsight-DLP-difference}
    We have
    \[\OPTstrong - O\left(\sqrt{T}\right) \leq \mathbb E_{\bm v, \bm p}\left[\totalRevenueHindsight\right] \leq \OPTstrong. \]  
    Meanwhile, when $\OPTstrong$ is a dual degenerate linear programming, we further have
    \[\OPTstrong - \mathbb E_{\bm v, \bm p}\left[\totalRevenueHindsight\right] = \Omega\left(\sqrt{T}\right). \]
\end{proposition}

Thus, by \Cref{prop:throttling-stochastic-lower-upper,,prop:pacing-stochastic-upper,,prop:hindsight-DLP-difference}, our problem reduces to a comparison between the two fluid benchmarks $\OPT$ and $\OPTstrong$. Now, a straight observation is that $\OPTstrong$ is no smaller than $\OPT$ with a larger feasible space. In fact, notice that both $\OPTstrong / T$ and $\OPT / T$ are constants since $\rho$ is a constant, it is common that $\OPTstrong - \OPT = \Theta(T)$. We next present precise sufficient conditions for the above to establish. 

To start with, we suppose that $G$ is a continuous distribution with density bounded from below. 
\begin{assumption}\label{assump:G-continuity}
    $G$ is a continuous distribution on $[0, \vmax]$ with density strictly no less than some constant $L > 0$. 
\end{assumption}

The second assumption restricts distribution $G$ from another side, which supposes that $r(v) / c(v)$ is not almost surely constant on $(0, \vmax]$. Note that $c(v) > 0$ when $v > 0$ under \Cref{assump:G-continuity}. Therefore $r(v) / c(v)$ is well-defined on $(0, \vmax]$. 
\begin{assumption}\label{assump:r/c-non-constant}
    For any $\lambda > 0$, the measure that $r(v) / c(v) \neq \lambda$ is positive concerning distribution $F$. 
\end{assumption}

Under these two assumptions, we have the following theorem, which is proved in \Cref{sec:proof-comparison}. 
\begin{theorem}\label{thm:linear-difference-throttling-pacing}
    Under \Cref{assump:G-continuity,assump:r/c-non-constant}, when $\mathbb E_{v, p} [p \mathbf 1[v \geq p]] > \rho$, we have
    \[\mathbb E_{\bm v, \bm p}\left[\totalRevenuePacing\right] - \mathbb E_{\bm v, \bm p, \bm \gamma}\left[\totalRevenueThrottling\right] = \Theta\left(T\right). \]
\end{theorem}

Now, we emphasize that \Cref{assump:G-continuity,assump:r/c-non-constant} are not strong assumptions by remarking that \Cref{assump:G-continuity} holds for most parameterized continuous distribution families, and \Cref{assump:r/c-non-constant} holds unless $G$ has a special form, e.g., $G$ is a uniform distribution on $[0, \vmax]$. Therefore, we obtain that under stochastic values, dynamic pacing \citep{DBLP:journals/mansci/BalseiroG19} would outperform even the optimal online throttling strategy by a linear term on the buyer's expected revenue in general. 

Nevertheless, for the completeness of this part, we also mention some special cases in which $\OPT = \OPTstrong$, thus, these two strategies have asymptotically similar performances. 
\begin{theorem}\label{thm:comparable-throttling-pacing}
    Let $\beta$ be the \algName algorithm. When (a) the highest competing bid $p$ is fixed, or (b) $\mathbb E_{v, p} [p \mathbf 1[v \geq p]] \leq \rho$, then under full or partial information feedback, we have
    \begin{align*}
        \left|\mathbb E_{\bm v, \bm p, \bm \gamma}\left[\totalRevenueThrottling\right] - \mathbb E_{\bm v, \bm p}\left[\totalRevenuePacing\right]\right| = O\left(\sqrt{T}\right), \\ 
        \left|\mathbb E_{\bm v, \bm p}\left[\totalRevenueGeneral\right] - \mathbb E_{\bm v, \bm p}\left[\totalRevenuePacing\right]\right| = O\left(\sqrt{T}\right). 
    \end{align*}
\end{theorem}

\subsection{Adversarial Values}

We take a similar analysis method for adversarial values. We first recall the result of \cite{DBLP:journals/mansci/BalseiroG19} on the performance of dynamic pacing in this scenario. 
\begin{proposition}[From \cite{DBLP:journals/mansci/BalseiroG19}]\label{prop:pacing-adversarial-upper}
    We have the following: 
    \[\liminf_{T \to \infty, B = \rho T} \inf_{\bm v, \bm p} \left(\frac{1}{T} \cdot \left(\totalRevenuePacing - \frac{\rho}{\vmax} \cdot \totalRevenueHindsight\right)\right) \geq 0. \]
\end{proposition}

A key point in \Cref{prop:pacing-adversarial-upper} is that the adaptive pacing algorithm can deal with the scenario when both $\bm v$ and $\bm p$ are adversarial and reaches optimality in this case \citep{DBLP:journals/mansci/BalseiroG19}. Nevertheless, as revealed by \Cref{thm:impossibility-result}, this is impossible for any throttling strategy. The main reason here is that for throttling, the buyer can only decide whether to enter or save in each round, therefore, could suffer from a possible inefficient win when choosing to enter or a regrettable loss when choosing to save. However, for pacing, the buyer can filter out those unsatisfying chances by controlling the bid, and only winning when the reward is high and the cost is low. Consequently, pacing brings high performance even when the highest competing bid is adversarial, while throttling cannot achieve the same effect. 

However, when only $\bm v$ is adversarial and $\bm p$ is stochastic, notice that \Cref{thm:upper-bound-adversarial} can be adapted to arbitrary bidding strategies (which certainly includes throttling and pacing) without any modification on the proof. Therefore, combining our positive result on the \algName algorithm, we conclude that the throttling strategy and \algName is asymptotically optimal under this scenario. 

\section{Concluding Remarks}

In this work, we comprehensively discuss the dynamic throttling strategy in repeated second-price auctions from a buyer's perspective. 
Specifically, we consider the best possible performance of any throttling algorithm when the buyer's values and the highest competing bids are stochastic or adversarial. 
We show an impossibility result with adversarial highest competing bids and propose an \algName algorithm which reaches (near) optimality under full or partial information structure regardless of the value input model when the highest competing bid is stochastic. 
We also compare the dynamic throttling strategy to dynamic pacing under different settings. 
With stochastic values, dynamic throttling generally faces a linear gap compared with dynamic pacing on buyer's revenue. 
However, with adversarial values, we show that dynamic throttling and \algName is asymptotically the best bidding strategy with full or partial information feedback.

\newpage


\bibliographystyle{plainnat}

\bibliography{reference}

\newpage
\appendix

\section{Missing Proofs in \Cref{sec:b-i-r}}\label{sec:proof-b-i-r}

\subsection{Proof of \Cref{thm:lower-bound-stochastic}}

We suppose that $\vmax = 1$ and $\rho = 1/2$ for convenience, and a similar instance can be constructed for other $(\rho, \vmax)$ pairs. We construct such a problem instance with $v$ always equal to 1, $G$ be the uniform binary distribution on $1/3$ and $2/3$. Consider the optimal strategy $\pi^*$ for $\OPT$. Under this instance, we have $\mathbb E_{p\sim G}\left[p\right] = 1/2$. Therefore, $\pi^*(1/3) = \pi^*(2/3) = 1$, and $\OPT = \theta^*\cdot \mathbb E_{p\sim G}\left[1 - p\right] = 1/2$. 

Now, we consider any online throttling strategy $\beta$ when $4 | T$. Since $\bm v$ always equals $1^T$, we simplify the notation of $\totalRevenueGamma$ to $\totalRevenueNoV$ in the context of the given instance. We give an upper bound on $\totalRevenueNoV$ in the following lemma, the proof of which can be found in \Cref{app:upper-bound-revenue-throttling}.

\begin{lemma}\label{lem:upper-bound-revenue-throttling}
    Given a realization of $\bm p \sim G^T$, we let $S$ be the number of $1/3$(s) in $\bm p$, then 
    \begin{gather*}
        \totalRevenueNoV \leq 
        \begin{cases}
            \frac{1}{3} \cdot (S + T) & \text{if } S \geq T / 2, \\
            \frac{2}{3} \cdot S + \frac{1}{3} \cdot  \left\lfloor \frac{3T - 2S}{4} \right\rfloor & \text{if } S < T/2. 
        \end{cases}
    \end{gather*}
\end{lemma}

With \Cref{lem:upper-bound-revenue-throttling}, we can come to analyze the lower bound of the regret of any online throttling strategy. We first give a non-simplified formula on the regret with the following lemma, the proof of which can be found in \Cref{app:regret-lower-bound-non-simplified}.
\begin{lemma}\label{lem:regret-lower-bound-non-simplified}
    In the given instance $(F, G)$, for any strategy $\beta$ and number of rounds $T$, we have
    \[\OPT - \mathbb E_{\bm p \sim G^T}\left[\totalRevenueNoV\right] \geq \frac{1}{12\cdot 2^T}\sum_{t = 1}^{T/4} \left(T - 4(t - 1)\right)\binom{T + 1}{2t - 1}. \]
\end{lemma}

Now, we need to simplify the lower bound we give above, which we finish in the following lemma. In fact, the complex summing formula turns out to be much more concise. 
\begin{lemma}\label{lem:lower-bound-simplified}
    \[\sum_{t = 1}^{T/4} \left(T - 4(t - 1)\right)\binom{T + 1}{2t - 1} = 2^{T - 1} + \frac{T}{2}\binom{T}{T/2}. \]
\end{lemma}
The proof of \Cref{lem:lower-bound-simplified} can be found in \Cref{app:lower-bound-simplified}.

At last, by combining \Cref{lem:regret-lower-bound-non-simplified} and \Cref{lem:lower-bound-simplified} and noticing that $\binom{T}{T/2} \geq 2^T / \sqrt{2T}$ by Stirling's formula, we have
\begin{align*}
    \mathbb E_{\bm p \sim G^T} \left[\regretNoV\right] &= \OPT - \mathbb E_{\bm p \sim G^T}\left[\totalRevenueNoV\right] \\
    &\geq \frac{1}{12\cdot 2^T}\left(2^{T - 1} + \frac{T}{2}\binom{T}{T/2}\right) \\
    &\geq \frac{1}{24} + \frac{\sqrt{2}}{48}\sqrt{T}. 
\end{align*}
 Therefore, \Cref{thm:lower-bound-stochastic} holds with $\constantLowerBound = \sqrt{2} / 48$. 

\subsection{Proof of \Cref{thm:upper-bound-adversarial}}

We prove the theorem in the following two steps. To start with, we employ Yao's principle \citep{yao1977probabilistic}
and shift our problem into bounding the performance of any \emph{deterministic} algorithms when $\bm p$ is stochastic and $\bm v$ follows a vector distribution. We then finish the proof by constructing a ``hard'' instance for any throttling strategy in this case. For brevity, for any $\mu$, we give the following definition: 
\[\discountedRegretGamma \coloneqq \frac{1}{T}\cdot \left[\totalRevenueGamma - \mu\cdot  \totalRevenueHindsight\right]. \]

\paragraph{Step 1: Applying Yao's principle.} We first give the following lemma, a variant of Yao's principle, which helps us to change our problem into an easier one. 
\begin{lemma}[A variant of Yao's principle]\label{lem:yao-principle}
    Let $\mathcal{V}$ denote any particular distribution of $\bm v$ support on $\{\bm v^1, \cdots, \bm v^m\} \subseteq [0, \vmax]^{T}$. Then for any distribution $G'$ with support $[0, \vmax]$ and any throttling strategy $\beta$, we have 
    \[
        \inf_{\bm v, G}\mathbb{E}_{\bm p\sim G^T, \bm \gamma} \left[\discountedRegretGamma\right] \leq \sup_{\bm \gamma}\mathbb{E}_{\bm p\sim (G')^T, \bm v\sim \mathcal{V}} \left[\discountedRegretGamma\right]. 
    \]
\end{lemma}

With \Cref{lem:yao-principle}, we are left to consider the optimal deterministic algorithm, which is seen as a random algorithm with the optimal random tape. Therefore, we omit the randomness $\bm \gamma$ in the following. 

\paragraph{Step 2: Construction of a ``hard'' instance.} 
We now construct a problem instance that traps any deterministic throttling strategy. We suppose $T\geq \vmax/\rho$. Let the distribution $G'$ specified in \cref{lem:yao-principle} be a one-point distribution on some $p < \vmax$. In other words, $p_t = p < \vmax$ for all $t \in [T]$ with the value of $p$ to be decided later. Therefore, we have
\[\bm p = (\underbrace{p, \cdots, p}_{T\text{ auctions}}). \]

We now split the sequence into $(m + 1)$ batches with $m$ a constant with $m > \lceil \vmax/\rho\rceil$. The starting $m$ batches are of length $\lfloor T/m\rfloor$ each, and the ending batch is of length $T' \coloneqq T - m\lfloor T/m\rfloor$.

We continue to construct the distribution $\mathcal V$ of $\bm v$ in \Cref{lem:yao-principle}. We let the support of $\mathcal V$ be the following vectors: 
\begin{align*}
    \bm v^1 &= (\underbrace{v_1, \cdots ,v_1}_{\lfloor T/m\rfloor\text{ auctions}}, \underbrace{v_2, \cdots, v_2}_{\lfloor T/m\rfloor\text{ auctions}}, \cdots, \underbrace{v_{m-1}, \cdots, v_{m-1}}_{\lfloor T/m\rfloor\text{ auctions}}, \underbrace{v_m, \cdots, v_m}_{\lfloor T/m\rfloor\text{ auctions}}, \underbrace{p, \cdots, p}_{T'\text{ auctions}}), \\
    \bm v^2 &= (\underbrace{v_1, \cdots ,v_1}_{\lfloor T/m\rfloor\text{ auctions}}, \underbrace{v_2, \cdots, v_2}_{\lfloor T/m\rfloor\text{ auctions}}, \cdots, \underbrace{v_{m-1}, \cdots, v_{m-1}}_{\lfloor T/m\rfloor\text{ auctions}}, \underbrace{p, \cdots\cdots, p}_{\lfloor T/m\rfloor\text{ auctions}}, \underbrace{p, \cdots, p}_{T'\text{ auctions}}), \\
    &\vdots\\
    \bm v^{m-1} &= (\underbrace{v_1, \cdots ,v_1}_{\lfloor T/m\rfloor\text{ auctions}}, \underbrace{v_2, \cdots, v_2}_{\lfloor T/m\rfloor\text{ auctions}}, \underbrace{p, \cdots\cdots\cdots\cdots\cdots\cdots\cdots\cdots\cdots\cdots p}_{(m-2)\lfloor T/m\rfloor\text{ auctions}}, \underbrace{p, \cdots, p}_{T'\text{ auctions}}), \\
    \bm v^{m} &= (\underbrace{v_1, \cdots ,v_1}_{\lfloor T/m\rfloor \text{ auctions}}, \underbrace{p, \cdots\cdots\cdots\cdots\cdots\cdots\cdots\cdots\cdots\cdots\cdots\cdots\cdots\cdots\cdot p}_{(m-1)\lfloor T/m\rfloor \text{ auctions}}, \underbrace{p, \cdots, p}_{T' \text{ auctions}}). \\
\end{align*}
Here, $v_j = p(1+\epsilon^{m+1-j})$, with exact value of $\epsilon\in (0,1]$ to be determined later. This notation should not be confused with the $v_t$ notation we use throughout this work. Since $v_m \leq \vmax$, we have $p \leq \vmax / (1 + \epsilon)$. Further, we let the probability that $\mathcal V$ takes $\bm v^i$ be $q_i (1 \leq i \leq m)$, where 
\[q_1 = \epsilon^{m - 1}, q_2 = \epsilon^{m - 2} - \epsilon^{m - 1}, \cdots, q_{m - 1} = \epsilon - \epsilon^2, q_m = 1 - \epsilon. \]

We now proceed by simplifying a buyer's deterministic throttling strategy. We start by considering the value sequence $\bm v^1$. First, notice that auctions are identical within a batch. Therefore, a deterministic strategy only needs to specify how many auctions to attend in each batch. Combining that the buyer never gets positive revenue in the final batch, any deterministic throttling strategy facing $\bm v^1$ degenerates to an integer vector $\bm x = (x_1, \cdots, x_m)$ such that 
\[0 \leq x_1, \cdots, x_m \leq \left\lfloor \frac{T}{m}\right\rfloor, x_1 + \cdots + x_m \leq \left\lfloor \frac{\rho T}{p}\right\rfloor. \]
The constraint is given by the budget since the payment in each auction is $p$ as long as the strategy participates. We now consider the above strategy facing other value sequences. Specifically, for any $i > 1$, we notice that $\bm v^i$ coincides with $\bm v^1$ at first $i \cdot \lfloor T / m \rfloor$ auctions. Consequently, these two sequences are indistinguishable for any strategy on this prefix. Meanwhile, once $\bm v^i$ deviates from $\bm v^1$, the strategy's choice does not affect the accumulated revenue anymore as the net revenue is $0$ for each auction in the suffix. In conclusion to this part of the discussion, we state that any deterministic throttling strategy is equivalent to the integer vector $\bm x$ specified above. 

We now let $u_i \coloneqq v_i - p = p\cdot \epsilon^{m + 1 - i}$ for $1 \leq i \leq m$, further define the utility matrix $U$ as: 
\begin{equation*}
    U \coloneqq \begin{bmatrix}
        u_1 & u_2 & \cdots & u_{m-1} & u_m \\
        u_1 & u_2 & \cdots & u_{m-1} & 0 \\
        \vdots & \vdots &  & \vdots & \vdots \\
        u_1 & u_2 & \cdots & 0 & 0 \\
        u_1 & 0 & \cdots & 0 & 0 \\
    \end{bmatrix}. 
\end{equation*}

We write $\bm q \coloneqq (q_1, \cdots, q_m)^\top$. Hence, for any deterministic throttling strategy $\beta$ that is equivalent to the vector $\bm x$ as stated above, we can calculate its expected total revenue as:
\begin{align}
    \mathbb E_{\bm p \sim (G')^T, \bm v \sim \mathcal V} \left[\totalRevenueGeneral\right] = \bm q^\top U \bm x = p\epsilon^m (x_1 + \cdots + x_m) \overset{(\text a)}{\leq} \epsilon^m \cdot \rho T. \label{eq:upper-bound-any-strategy}
\end{align}
Here, (a) is by the restriction on vector $\bm x$. Meanwhile, we give a lower bound on the hindsight benchmark. Due to $m = \lceil \vmax / \rho \rceil$ and $p < \vmax$, we know that $p \cdot \lfloor T / m \rfloor \leq \rho T = B$. In other words, for each $\bm v^i$, the hindsight value $\totalRevenueHindsight$ wins at least $\lfloor T / m \rfloor$ auctions each with revenue $u_{m + 1 - i} = p\epsilon^i$. Thus, we have
\begin{align}
    \mathbb E_{\bm p \sim (G')^T, \bm v \sim \mathcal V} \left[\totalRevenueHindsight\right] &\geq \sum_{i = 1}^m q_i\left\lfloor \frac{T}{m} \right\rfloor p \epsilon^i \notag \\
    &= \left\lfloor \frac{T}{m} \right\rfloor p \cdot \epsilon^m (m - \epsilon(m - 1)) \notag \\
    &> \left(\frac{T}{m} - 1\right) p \cdot \epsilon^m (m - \epsilon(m - 1)). \label{eq:lower-bound-hindsight-optimum}
\end{align}

Now, when $\mu > \rho / \vmax$, suppose $\mu =  (1 - \delta)^{-1}\cdot (\rho / \vmax)$ for some $\delta \in (0, 1)$. Combining \eqref{eq:upper-bound-any-strategy} and \eqref{eq:lower-bound-hindsight-optimum} together, we take $p = \vmax / (1 + \epsilon)$ and $\epsilon = \delta / (4 - 2\delta)$ to obtain that
\begin{align*}
    \mathbb E_{\bm p \sim (G')^T, \bm v \sim \mathcal V}\left[\discountedRegret\right] &< \frac{\epsilon^m \rho}{1 - \delta} \cdot \left(1 - \delta - \frac{p}{\vmax} (m - \epsilon (m - 1))\cdot\left(\frac{1}{m} - \frac{1}{T}\right)\right) \\
    &\overset{(\text a)}{=} \frac{\epsilon^m \rho}{(1 - \delta)(1 + \epsilon)} \cdot \left((1 - \delta)(1 + \epsilon) - (m - \epsilon (m - 1)) \cdot\left(\frac{1}{m} - \frac{1}{T}\right)\right) \\
    &= \frac{\epsilon^m \rho}{(1-\delta)(1+\epsilon)} \cdot \left(-\delta+\left(1-\delta+\frac{m-1}{m}\right)\epsilon + \frac{m}{T} - \frac{\epsilon(m-1)}{T}\right)\\
    &\leq \frac{\epsilon^m \rho}{(1-\delta)(1+\epsilon)} \cdot \left(-\delta+\left(2-\delta\right)\epsilon + \frac{m}{T}\right)\\
    &\overset{(\text b)}{=} \frac{\epsilon^m \rho}{(1-\delta)(1+\epsilon)} \cdot \left(-\frac{\delta}{2} + \frac{m}{T}\right).   
\end{align*}
In the above, (a) and (b) hold correspondingly due to the value of $p$ and $\epsilon$. Therefore, when $T > 4m / \delta$, we derive that $\mathbb E_{\bm p \sim (G')^T, \bm v \sim \mathcal V}[\discountedRegret] < 0$. In other words, combining with \Cref{lem:yao-principle}, we conclude that when $\mu > \vmax / \rho$, 
\begin{align*}
    \liminf_{T \to \infty, B = \rho T} \inf_{\bm v, G}\mathbb{E}_{\bm p\sim G^T, \bm \gamma} \left[\discountedRegretGamma\right] \leq \liminf_{T \to \infty, B = \rho T} \sup_{\bm \gamma}\mathbb{E}_{\bm p\sim (G')^T, \bm v\sim \mathcal{V}} \left[\discountedRegretGamma\right] < 0. 
\end{align*}
This finishes the proof of the theorem.

\subsection{Proof of \Cref{thm:impossibility-result}}

In the proof, we suppose $\vmax = 1$ and $\rho = 1/3$ without loss of generality. In this case, consider the condition where $v_t=2/3$ for any $t\in [T]$. $p_t$ will be given dynamically as the auction proceeds. In our construction, if $x_t=1$, let $p_t$ be $2/3-\epsilon$, and if $x_t=0$, then let $p_t$ be $\epsilon$.
The exact value of $\epsilon < 1/3$ will be given below.

We then can derive that the number of $t$ satisfying $x_t=1$ must be less than $T/(2-3\epsilon)$.
Therefore, we obtain 
\[
\totalRevenueGamma \leq \frac{\epsilon}{2-3\epsilon}T.
\]
However, for $\totalRevenueHindsight$ which knows the trajectory $\bm p$, we can take a solution such that $x_t = 1$ when $p_t=\epsilon$ and $x_t = 0$ otherwise. Since $\epsilon < 1/3$, such a solution does not violate the budget constraint. Note that there are at least $T - T/(2 - 3\epsilon) = (1 - 3\epsilon) T / (2 - 3\epsilon)$ times that $p_t = \epsilon$. we achieve that 
\[\totalRevenueHindsight\geq \frac{1-3\epsilon}{2-3\epsilon}\cdot T \cdot\frac{2-3\epsilon}{3}=\frac{1-3\epsilon}{3} T.
\]
Now we take $\epsilon= \mu / (3 \mu + 6) < 1/3$ and get that 
\[
\frac{\totalRevenueGamma}{\totalRevenueHindsight} \leq \frac{3\epsilon}{(1-3\epsilon)(2-3\epsilon)} \leq \frac{3\epsilon}{1-3\epsilon} = \frac{\mu}{2}.
\]
As a result, we obtain that 
\[\frac{1}{T}\left(\totalRevenueGamma - \mu\cdot \totalRevenueHindsight\right) \leq -\frac{\mu}{2}\cdot \frac{\totalRevenueHindsight}{T} \leq - \frac{\mu}{3 \mu + 6} < 0. \]
This concludes the proof.

\section{Missing Proofs in \Cref{sec:algorithm-performance}}
\label{sec:proof-algo}

\subsection{Proof of \Cref{lem:estimation-r-c}}

In the full-information setting, the buyer can observe all the history prices (\Cref{algoline:observe_condition}), i.e., for any $t > 1$, $|\mathcal I_t| = \{1, 2, \cdots, t-1\}$.  Note that \Cref{algoline:estimate} of \algName essentially estimates $r(v_t)$ and $c(v_t)$ according to an empirical distribution $\estG_t$. 
In detail, we let $\estG_t$ be the distribution of $(t - 1)$ independent samples of $p$, then we have
\begin{gather*}
    \estr_t(v_t) = \frac{1}{t - 1}\sum_{1 \leq \tau \leq t - 1} (v_t - p_\tau)^+ + \epsilon_tv_t = \mathbb E_{p \sim \estG_t} \left[(v_t - p)^+\right] + \epsilon_tv_t, \\
    \estc_t(v_t) = \frac{1}{t - 1}\sum_{1 \leq \tau \leq t - 1} p_\tau\mathbf 1[v_t \geq p_\tau] - 2\epsilon_tv_t = \mathbb E_{p \sim \estG_t} \left[p\mathbf 1[v_t \geq p]\right] - 2\epsilon_tv_t. 
\end{gather*}
We now do a calculation on $r(v_t)$ and $c(v_t)$. By using integration by parts, we have
\begin{align*}
    r(v_t) = \mathbb E_{p\sim G}\left[(v_t - p)^+\right] & = \int_{0}^{v_t} (v_t - p)\diffe G(p) = - \int_{0}^{v_t} G(p)\diffe (v_t - p) = \int_{0}^{v_t} G(p)\diffe p, \\
    c(v_t) = \mathbb E_{p\sim G}\left[p\mathbf 1[v_t \geq p]\right] & = \int_{0}^{v_t} p\diffe G(p) = v_tG(v_t) - \int_{0}^{v_t} G(p)\diffe p. 
\end{align*}

By DKW inequality, for any $\epsilon_t \geq 0$, we have
\[\Pr\left[\sup_{p \in [0, \vmax]}\left|\estG_t(p) - G(p)\right| \geq \epsilon_t\right] \leq 2\exp \left(-2(t - 1)\epsilon_t^2\right) = q(t, \epsilon_t). \]
Suppose that $\left|\estG_t(p) - G(p)\right| \leq \epsilon_t$ happens for every $p \in [0, \vmax]$, 
then for any $v_t \in [0, \vmax]$, we have
\begin{gather*}
    \left|\mathbb E_{p\sim \estG_t}\left[(v_t - p)^+\right] - \mathbb E_{p\sim G}\left[(v_t - p)^+\right]\right| 
    = \left|\int_{0}^{v_t} \left(\estG_t(p) - G(p)\right)\diffe p\right| 
    \leq \epsilon_tv_t, \\
    \left|\mathbb E_{p\sim \estG_t}\left[p\mathbf 1[v_t \geq p]\right] - \mathbb E_{p\sim G}\left[p\mathbf 1[v_t \geq p]\right]\right| 
    = \left|v_t\left(\estG_t(v_t) - G(v_t)\right) - \int_{0}^{v_t} \left(\estG_t(p) - G(p)\right)\diffe p\right|
    \leq 2\epsilon_tv_t. 
\end{gather*}

As a result, by the definition of $\estr_t(\cdot)$ and $\estc_t(\cdot)$ in \Cref{alg:OGD-CB}, we derive the lemma. 

\subsection{Proof of \Cref{lem:lambda_upper_bound}}

First, observe that 
\begin{align*}
    \estr_t(v_t) & = \mathbb E_{p \sim \estG_t} \left[(v_t - p)^+\right] + \epsilon_tv_t \\
    & = \mathbb E_{p \sim \estG_t} \left[v_t\mathbf 1\left[v_t \geq p_t\right]\right] - \mathbb E_{p \sim \estG_t} \left[p_t\mathbf 1\left[v_t \geq p_t\right]\right] + \epsilon_tv_t \\
    & = \mathbb E_{p \sim \estG_t} \left[v_t\mathbf 1\left[v_t \geq p_t\right]\right] - \estc_t(v_t) - \epsilon_tv_t.
\end{align*}
Then by the choice of $x_t$, we have 
\begin{equation}\label{eqn:x_tc_t_upper_bound}
    x_t\estc_t(v_t) \leq \frac{x_t}{1 + \lambda_t} \left(\mathbb E_{p_t\sim \estG_t}\left[v_t\mathbf 1\left[v_t \geq p_t\right]\right] - \epsilon_tv_t \right) \leq \frac{\vmax}{1 + \lambda_t},
\end{equation}
since the value of $x_t (\estr_t(v_t) - \lambda_t\estc_t(v_t))$ must be no less than 0. If we assume $\lambda_t \leq {\vmax}/{\rho} - 1$, we have
\begin{align*}
    \lambda_{t+1} &= (\lambda_{t} + \eta_t (x_t\estc_t(v_t) - \rho))^+ \overset{(\text a)}{\leq} \left(\lambda_{t} + \frac{\eta_t \vmax}{1 + \lambda_t} - \eta_t \rho \right)^+ \\ &\overset{(\text b)}{\leq} \max\left\{\frac{\vmax}{\rho} - 1, 0\right\} \overset{(\text c)}{=} \frac{\vmax}{\rho} - 1.
\end{align*}
In the derivation above, (a) follows the bound of $x_t \widetilde c_t(v_t)$ in \eqref{eqn:x_tc_t_upper_bound}; (b) is due to the non-decreasing of $\lambda_t + {\eta_t}v_t/{(1 + \lambda_t)}$ when $\lambda_t \in [0, \vmax/\rho-1]$ since $\eta_t v_t \leq 1$; (c) uses the fact that $\eta_t \leq 1 / \rho$ and $\rho \leq \vmax$. Since $\lambda_1 = \lambda_2 = 0 \leq \vmax / {\rho} - 1$, by induction we obtain that $\lambda_t \leq \vmax/{\rho}-1$ uniformly for all $t \leq T_0 + 1$.

\subsection{Proof of \Cref{lem:OGD}}

By the update rule of $\lambda_t$, we have for any $\lambda \in [0, \vmax/\rho - 1]$, 
\begin{align*}
    \|\lambda_{t+1} - \lambda\|^2 & \leq \|\lambda_{t} + \eta_t (x_t\estc_t(v_t) - \rho) - \lambda\|^2, 
\end{align*}
which implies that
\begin{align*}
    \frac{1}{\eta_t}\left(\|\lambda_{t+1} - \lambda\|^2 - \|\lambda_{t} - \lambda\|^2\right) & \leq (\lambda_t - \lambda)(x_t\estc_t(v_t) - \rho) + \eta_t\|(x_t\estc_t(v_t) - \rho\|^2.
\end{align*}
A telescoping summation from $t=2$ through $T_0$ gives
\begin{align*}
    \sum_{t=2}^{T_0} (\lambda_t - \lambda)(x_t\estc_t(v_t) - \rho) & \geq \sum_{t=2}^{T_0} \frac{1}{\eta_t}\left(\|\lambda_{t+1} - \lambda\|^2 - \|\lambda_{t} - \lambda\|^2\right) - \sum_{t=2}^{T_0} \eta_t\|(x_t\estc_t(v_t) - \rho\|^2 \\
    & \overset{(\text a)}{\geq} \sum_{t=2}^{T_0} \frac{1}{\eta_t}\left(\|\lambda_{t+1} - \lambda\|^2 - \|\lambda_{t} - \lambda\|^2\right) - \vmax^2 \sum_{t=2}^{T_0} \eta_t \\
    & \overset{(\text b)}{\geq} - \left(\frac{\vmax}{\rho} - 1\right)^2\sum_{t=3}^{T_0}\left(\frac{1}{\eta_t} - \frac{1}{\eta_{t-1}}\right) - \vmax^2 \sum_{t=2}^{T_0} \eta_t \\
    & \geq - \frac{\left(\vmax/\rho - 1\right)^2}{\eta_{T_0}} - \vmax^2 \sum_{t=2}^{T_0} \eta_t,
\end{align*}
where (a) holds since $|x_t\estc_t(v_t) - \rho| \leq \vmax$ and (b) follows from a rearrangement, \Cref{lem:lambda_upper_bound}, and that $\eta_t$ decreases in $t$.

\subsection{Proof of \Cref{thm:full-info-stochastic}}

We prove the result in four steps. We first lower bound the performance of our throttling strategy by using Azuma--Hoeffding inequality and \Cref{lem:estimation-r-c}. Then we rewrite the lower bound into two parts and bound each part respectively. The third step proves that the ending time $T_0$ of our strategy is close to $T$. Finally, we conclude by combining these steps to obtain the desired result.

\paragraph{Step 1: lower bound on the performance.} We start with lower-bounding the total reward that the algorithm gets. Notice that 
\[\mathbb E_{v_t \sim F, p_t\sim G}\left[x_t(r_t - r(v_t)) \mid \mathcal H_t\right] = 0. \]
Thus, by applying Azuma--Hoeffding inequality, we have
\begin{equation}\label{eqn:xtvt_step_1a}
    \totalRevenueGeneral = \sum_{t = 1}^{T_0} x_tr_t\geq \sum_{t=1}^{T_0}x_tr(v_{t}) - \vmax\sqrt{2T \log T}, 
\end{equation}
with a failure probability of $1/T$. Suppose that the event stated in \Cref{lem:estimation-r-c} happens, i.e. all the estimations are successful, which holds with probability at least $1 - \sum_{t=2}^{T_0} q(t, \epsilon_t)$. In this case, \eqref{eq:r-estimation} establishes for all $2 \leq t \leq T_0$ and hence
\begin{equation}\label{eqn:xtvt_step_1b}
    \sum_{t=2}^{T_0}x_tr(v_t)\geq \sum_{t=2}^{T_0}x_t\estr_t(v_t)-2\sum_{t=2}^{T_0}x_t\epsilon_tv_t. 
\end{equation}
Here, we start the summation from $t = 2$ since $\estr_t(\cdot)$ is not defined for $t = 1$. By the construction of $\epsilon_t = \sqrt{(\log 2 + 2\log T) / (2(t - 1))}$ with full-information feedback, we have 
\begin{equation}\label{eqn:xtvt_step_1c}
    \sum_{t=2}^{T_0} \epsilon_t \leq \sqrt{2T(\log 2 + 2\log T)} \text{ , and } \sum_{t=2}^{T_0} q(t, \epsilon_t) \leq 1/T.
\end{equation}
Combining equations \eqref{eqn:xtvt_step_1a}, \eqref{eqn:xtvt_step_1b} and \eqref{eqn:xtvt_step_1c}, with probability at least $1 - 2/T$, we have
\begin{equation}\label{eqn:xtvt_step_1}
    \totalRevenueGeneral \geq \sum_{t=2}^{T_0}x_t\estr_t(v_{t}) - \left(\sqrt{2T \log T} + 2 \sqrt{2T(\log 2 + 2\log T)}\right)\cdot \vmax. 
\end{equation}

\paragraph{Step 2: bounding $\sum_{t=2}^{T_0}x_t\estr_t(v_{t})$.} We next consider the first term in the right hand side of \eqref{eqn:xtvt_step_1}.
Rewriting it into two parts, we have
\begin{equation}\label{eqn:xtvt_step_2a}
    \sum_{t=2}^{T_0}x_t\estr_t(v_{t}) = \underbrace{\sum_{t=2}^{T_0} \left(x_t\estr_t(v_t) - \lambda_t \left(x_t\estc_t(v_t) - \rho\right)\right)}_{R_1} + \underbrace{\sum_{t=2}^{T_0} \lambda_t\left(x_t\estc_t(v_t) - \rho\right)}_{R_2}.
\end{equation}

Let $\pi^*: [0, 1] \mapsto [0, 1]$ be the optimal solution for the benchmark $\OPT$ defined in Program \eqref{eq:OPT}. We have \begin{align}
    R_1 &= \sum_{t=2}^{T_0} x_t(\estr_t(v_t) - \lambda_t\estc_t(v_t)) + \sum_{t=2}^{T_0} \lambda_t\rho \overset{(\text a)}{\geq} \sum_{t=2}^{T_0} \pi^*(v_t)(\estr_t(v_t) - \lambda_t\estc_t(v_t)) + \sum_{t=2}^{T_0} \lambda_t \rho\notag\\
    &\overset{(\text b)}{\geq} \sum_{t=2}^{T_0} \pi^*(v_t)(r(v_t) - \lambda_tc(v_t)) + \sum_{t=2}^{T_0} \lambda_t\rho = \sum_{t=2}^{T_0} \left(\pi^*(v_t)(r(v_t) - \lambda_tc(v_t)) + \lambda_t\rho\right). \label{eqn:xtvt_step_2b}
\end{align}
Here (a) holds by the decision rule in \Cref{algoline:action}; (b) follows from \Cref{lem:estimation-r-c}. Note that (b) brings no additional probability of failure as we already assume \eqref{eq:r-estimation} and \eqref{eq:c-estimation} holds for all $t$ in the first step.

As $\pi^*$ is optimal and feasible, we have
\begin{align}
    T \cdot \mathbb E_{v_t \sim F} [\pi^*(v_t) r(v_t)] & = \OPT, \label{eqn:xtvt_step_2c}\\
    \mathbb E_{v_t \sim F} [\pi^*(v_t) c(v_t)] & \leq \rho. \label{eqn:xtvt_step_2d}
\end{align}
Notice that $\lambda_t$ and $v_t$ are independent, so we further have for any $t$,
\begin{equation}\label{eqn:xtvt_step_2e}
    \mathbb E_{v_t \sim F} \left[\lambda_t \pi^*(v_t) c(v_t) - \lambda_t \rho \mid \mathcal H_t\right] \leq 0.
\end{equation}
Combining \eqref{eqn:xtvt_step_2c} and \eqref{eqn:xtvt_step_2e}, we have
\[\mathbb E_{v_t \sim F}\left[\pi^*(v_t)(r(v_t) - \lambda_tc(v_t)) - \left(\frac{\OPT}{T} - \lambda_t\rho \right) \,\middle\vert\, \mathcal H_t\right] \geq 0. \]
Thus, we can again apply Azuma–Hoeffding inequality and show that with failure probability $O(1/T)$, 
\begin{equation}\label{eqn:xtvt_step_2f}
    R_1 \geq \sum_{t=2}^{T_0} \left(\pi^*(v_t)(r(v_t) - \lambda_tc(v_t)) + \lambda_t\rho\right) \geq \frac{T_0 - 1}{T} \OPT - \vmax\sqrt{2T \log T}.
\end{equation}

As for the other part $R_2$, with the help of \Cref{lem:OGD}, we have
\begin{equation}\label{eqn:xtvt_step_2g}
    R_2 \overset{(\text a)}{\geq} - \left(\frac{\left(\vmax/\rho - 1\right)^2}{\eta_{T_0}} + \sum_{t=2}^{T_0} \eta_t\right) \overset{(\text b)}{\geq} - \left((\vmax / \rho - 1)^2  + 2\right)\vmax\sqrt{T},
\end{equation}
where (a) holds by setting $\lambda=0$ in \eqref{eqn:OGD} and (b) holds since $\eta_t=(1/\rho - 1) / \sqrt{t}$ and $T_0 \leq T$. 

Applying \eqref{eqn:xtvt_step_2f} and \eqref{eqn:xtvt_step_2g} to \eqref{eqn:xtvt_step_2a}, we have
\begin{equation}\label{eqn:xtvt_step_2}
    \sum_{t=2}^{T_0}x_t\estr_t(v_{t}) \geq \frac{T_0 - 1}{T} \OPT - \vmax \left(\sqrt{2T \log T} + \left((\vmax / \rho - 1)^2 + 2\right)\sqrt{T}\right), 
\end{equation}
which holds with an extra failure probability $1/T$. 

\paragraph{Step 3: bounding the stopping time $T_0$.} This step aims to show that \Cref{alg:OGD-CB} does not run out of budget too early so that $T_0$ is actually close to $T$ with high probability. Intuitively, we prove this by arguing that the expenditure rate of \Cref{alg:OGD-CB} is fairly close to the target rate $\rho$. By the update rule we have for every $t \in \{2, \cdots, T_0\}$, $$\lambda_{t+1} \geq \lambda_{t} + \eta_t (x_t\estc_t(v_t) - \rho).$$ Reordering terms and summing up from $2$ through $T_0$, one has
\begin{align} \label{eqn:update_telescope}
    \sum_{t=2}^{T_0} \left(x_t\estc_t(v_t) - \rho\right) & \leq \sum_{t=2}^{T_0} \frac{\lambda_{t+1} - \lambda_t}{\eta_t} \overset{(\text a)}{=} \frac{\lambda_{T_0+1}}{\eta_{T_0}} + \sum_{t=3}^{T_0} \left(\frac{1}{\eta_{t-1}} - \frac{1}{\eta_{t}}\right)\lambda_t \notag \\ 
    & \overset{(\text b)}{\leq} \frac{\lambda_{T_0+1}}{\eta_T}  \overset{(\text c)}{\leq} \frac{\vmax/\rho - 1}{\eta_{T}}. 
\end{align}
where (a) holds since $\lambda_2 = 0$; (b) uses the uses the construction that $\eta_t$ is decreasing; and (c) holds by \Cref{lem:lambda_upper_bound}. Consequently, we have
\begin{align}\label{eq:boundt0}
    \rho(T - T_0 + 1) &\overset{(\text a)}{\leq} \sum_{t=2}^{T_0} x_tc_t + 2\vmax - \rho\left(T_0 - 1\right) \notag\\
    & \overset{(\text b)}{\leq} \sum_{t=2}^{T_0} x_t \estc_t(v_t) - \rho\left(T_0 - 1\right) + \vmax\sqrt{2T \log T} +  4\sum_{t=2}^{T_0} x_t\epsilon_tv_t + 2\vmax\notag\\
    & \overset{(\text c)}{\leq} \sum_{t=2}^{T_0} \left(x_t \estc_t(v_t) - \rho\right) + \vmax \left(\sqrt{2T \log T} +  4\sqrt{2T(\log 2 + 2\log T)} + 2\right) \notag \\
    &\overset{(\text d)}{\leq} \vmax \left((\vmax / \rho - 1)\sqrt{T} + \sqrt{2T \log T} +  4\sqrt{2T(\log 2 + 2\log T)} + 2\right).
\end{align}
Here, (a) holds since $\sum_{t=2}^{T_0} x_tc_t + 2\vmax \geq B = \rho T$; (b) is derived by applying the inequality \eqref{eq:c-estimation} and Azuma--Hoeffding inequality to bound the difference between $\sum_{t=2}^{T_0} x_tc_t$ and $\sum_{t=2}^{T_0} x_t\estc_t(v_t)$; (c) follows from \eqref{eqn:xtvt_step_1c}; and (d) follows from \eqref{eqn:update_telescope} and $\eta_T = 1 / (\vmax \sqrt{T})$. Note that only the Azuma--Hoeffding inequality in (b) brings an extra failure probability of $1/T$.

Simple rearrangements establish that, 
\begin{equation}\label{eqn:xtvt_step_3}
    \frac{T_0 - 1}{T}\OPT \geq \OPT - \frac{\OPT}{T}\cdot \frac{\vmax}{\rho} \left((\vmax / \rho - 1)\sqrt{T} + \sqrt{2T \log T} +  4\sqrt{2T(\log 2 + 2\log T)} + 2\right).
\end{equation}

\paragraph{Step 4: Putting everything together.} Observe that $\OPT / T = \Theta(1)$ and $\rho = \Theta(1)$. Therefore, combining inequalities \eqref{eqn:xtvt_step_1}, \eqref{eqn:xtvt_step_2} and \eqref{eqn:xtvt_step_3}, there is a constant $\constantFull$ such that with probability at least $1-4/T$,
\[ 
\totalRevenueGeneral \geq \opt - \constantFull \sqrt{T \log T},
\]
which concludes the proof of \Cref{thm:full-info-stochastic}.

\subsection{Proof of \Cref{thm:full-info-adversarial}}

We prove the theorem in five steps. 
We first roughly bound the performance of \Cref{alg:OGD-CB} concerning estimation errors and the OCO regret. 
Then we settle these two parts correspondingly to achieve the theorem. 

\paragraph{Step 1: Bounding the accumulated rewards with estimation error.}
We first show that
\begin{equation}\label{eq:niceob}
\rho\cdot \estr_t(v_{t}) \leq \left(x_{t}(\estr_t(v_{t})-\lambda_{t} \estc_t(v_{t})) + \rho\cdot \lambda_{t}\right)\cdot \vmax.
\end{equation}
Note that when $x_t=0$, \eqref{eq:niceob} naturally holds because $\estr_t(v_t)\leq \lambda_t\estc_t(v_t)\leq \lambda_t \vmax$.  When $x_t=1$, we obtain
\begin{align*}
    \rho\cdot \estr_t(v_t) &= \vmax \cdot \estr_t(v_t)+\left(\rho - \vmax\right)\estr_t(v_t) \\
    &\overset{(\text a)}{\leq} \vmax\cdot \estr_t(v_t)+\left(\rho - \vmax\right)\lambda_t\estc_t(v_t) \\
    &=\left(\estr_t(v_{t})-\lambda_{t} \estc_t(v_{t})\right)\cdot \vmax+\rho\cdot \lambda_t\estc_t(v_t)\\
    &\overset{(\text b)}{\leq} \left(\estr_t(v_{t})-\lambda_{t} \estc_t(v_{t})+ \rho\cdot \lambda_t\right)\cdot \vmax,
\end{align*}
where (a) holds because $\estr_t(v_{t})-\lambda_{t} \estc_t(v_{t})\geq 0$ and $\rho\leq 1$, and (b) follows from $\estc_t(v_t)\leq 1$. Combining the two cases above, we finish the proof of \eqref{eq:niceob}.
Reordering terms and summing over $t=2,\cdots,T_0$ implies
\begin{equation}\label{eq:sumres}
   \vmax\cdot \sum_{t=2}^{T_0} x_t\estr_t(v_t)\geq  \rho \sum_{t=2}^{T_0}\estr_t(v_t) + \vmax\cdot \sum_{t=2}^{T_0}\lambda_t(x_t\estc_t(v_t)-\rho).
\end{equation}
Notice that $\estr_t(v_t)$ and $\estc_t(v_t)$ are related to the specific $\bm p$. We then take the expectation of \eqref{eq:sumres} and derive
\begin{equation}\label{eq:sumres-expect}
    \mathbb{E}_{\bm p\sim G^T}\left[\sum_{t=2}^{T_0} x_t\estr_t(v_t)\right]\geq  \frac{\rho}{\vmax}\cdot \mathbb{E}_{\bm p\sim G^T} \left[\sum_{t=2}^{T_0}\estr_t(v_t)\right]+ \mathbb{E}_{\bm p\sim G^T}\left[\sum_{t=2}^{T_0}\lambda_t(x_t\estc_t(v_t)-\rho)\right].
\end{equation}
The left side of \eqref{eq:sumres-expect} estimates $\totalRevenueGeneral$. Also, the first part of the right side is a near estimated upper bound of $\rho\cdot \totalRevenueHindsight$ regardless of the difference between $T_0$ and $T$, and the second part gives the regret of the online gradient descent procedure. 
We now give a closer look at these terms. 

\paragraph{Step 2: Bounding $\sum_{t=2}^{T_0}\lambda_t(x_t\estc_t(v_t)-\rho)$.}
For the last term in \eqref{eq:sumres-expect}, Since $\bm p$ is stochastic, \Cref{lem:OGD} still holds here, which implies that for any $\bm p$, 
\begin{equation}\label{eq:est-OGD0}
    \sum_{t=2}^{T_0}\lambda_t(x_t\estc_t(v_t)-\rho)\overset{(\text a)}{\geq} -\left(\frac{\left(\vmax/\rho - 1\right)^2}{\eta_{T_0}} + \sum_{t=2}^{T_0} \eta_t\right)\overset{(\text b)}{\geq} -\left((\vmax / \rho - 1)^2 \vmax + 2 / \vmax\right)\sqrt{T}. 
\end{equation}
Here, (a) holds by setting $\lambda=0$ in \eqref{eqn:OGD} and (b) holds since $\eta_t=1 / (\vmax\sqrt{t})$ and $T_0 \leq T$. 
 Taking expectation of \eqref{eq:est-OGD0} over $G^T$ implies
\begin{equation}\label{eq:est-OGD}
  \mathbb{E}_{\bm p\sim G^T}\left[\sum_{t=2}^{T_0}\lambda_t(x_t\estc_t(v_t)-\rho)\right]\geq -\left((\vmax / \rho - 1)^2 + 2\right)\vmax\sqrt{T}.
\end{equation}

\paragraph{Step 3: Bounding $\sum_{t=2}^{T_0}x_t\estr_t(v_t)$ and $\sum_{t=2}^{T_0}\estr_t(v_t)$.}
In this step, we apply \Cref{lem:estimation-r-c}. 
Suppose that the events stated in it happen for all $t$, i.e., the error of each estimation is bounded, which holds with probability at least $1 - \sum_{t=2}^{T_0} q(t, \epsilon_t)$. In this case, \eqref{eq:r-estimation} establishes for all $2 \leq t \leq T_0$. Hence, we obtain 
\begin{equation}\label{eq:est-epsilon1}
\sum_{t=2}^{T_0}x_t\estr_t(v_t)\leq \sum_{t=2}^{T_0}x_tr(v_t)+2\sum_{t=2}^{T_0}x_t\epsilon_tv_t
\end{equation}
and
\begin{equation}\label{eq:est-epsilon2}
     \sum_{t=2}^{T_0}\estr_t(v_t)\geq \sum_{t=2}^{T_0}r(v_t).
\end{equation}
By the construction of $\epsilon_t = \sqrt{(\log 2 + 2\log T) / (2(t - 1))}$, we have 
\begin{equation}\label{eq:est-epsilon3}
    \sum_{t=2}^{T_0} \epsilon_t \leq \sqrt{2T(\log 2 + 2\log T)} \text{ , and } \sum_{t=2}^{T_0} q(t, \epsilon_t) \leq 1/T.
\end{equation}

Combining \eqref{eq:est-epsilon1}, \eqref{eq:est-epsilon2} and \eqref{eq:est-epsilon3}, we obtain that with probability at least $1-1/T$, 
\begin{equation}\label{eq:est-epsilon4}
   \sum_{t=2}^{T_0}x_t\estr_t(v_t)\leq \sum_{t=2}^{T_0}x_tr(v_t)+2\vmax\sqrt{2T(\log 2 + 2\log T)} 
\end{equation}
and
\begin{equation}\label{eq:est-epsilon5}
   \sum_{t=2}^{T_0}\estr_t(v_t)\geq \sum_{t=2}^{T_0}r(v_t).
\end{equation}

Taking the expectation of \eqref{eq:est-epsilon4} over $G$ implies that
\begin{align}\label{eq:est-epsilon6}
   \mathbb{E}_{\bm p\sim G^T}\left[\sum_{t=2}^{T_0}x_t\estr_t(v_t)\right]&\leq \mathbb{E}_{\bm p\sim G^T}\left[\sum_{t=2}^{T_0}x_tr(v_t)\right]+\vmax\left(2\sqrt{2T(\log 2 + 2\log T)}+\frac{T_0}{T}\right)\notag\\
   &=\mathbb{E}_{\bm p\sim G^T}\left[\totalRevenueGeneral\right]-{r(v_1)}+\vmax\left(2\sqrt{2T(\log 2 + 2\log T)}+\frac{T_0}{T}\right). 
\end{align}
Similarly, for \eqref{eq:est-epsilon5}, we have 
\begin{align}
    \mathbb{E}_{\bm p\sim G^T}\left[\sum_{t=2}^{T_0}\estr_t(v_t)\right]&\geq\sum_{t=2}^{T_0}r(v_t)-\frac{T_0}{T}\cdot \vmax \notag \\
    &\geq \mathbb{E}_{\bm p\sim G^T}\left[\totalRevenueHindsight\right]-r(v_1)-\vmax\left(T-T_0+1 + \frac{T_0}{T}\right). \label{eq:est-epsilon7}
\end{align}

\paragraph{Step 4: Bounding $T_0$.}
Note that the estimation of $T_0$ in the proof of \Cref{thm:full-info-stochastic} still holds. Therefore, we multiply \eqref{eq:boundt0} by $1/\rho$ and achieve that 
\begin{equation}\label{eq:est-t0}
    T-T_0+1\leq \frac{\vmax}{\rho}\left((\vmax / \rho - 1)\sqrt{T} + \sqrt{2T \log T} +  4\sqrt{2T(\log 2 + 2\log T)} + 2\right).
\end{equation}

\paragraph{Step 5: Putting everything together.}
Finally, taking \eqref{eq:est-OGD}, \eqref{eq:est-epsilon6}, \eqref{eq:est-epsilon7} and \eqref{eq:est-t0} into \eqref{eq:sumres-expect}, we arrive at
\begin{align*}
\mathbb{E}_{\bm p\sim G^T}\left[\totalRevenueGeneral\right]&\geq \frac{\rho}{\vmax} \cdot \mathbb{E}_{\bm p\sim G^T}\left[\totalRevenueHindsight\right]-O\left(\sqrt{T\log T}\right).
\end{align*}
This concludes the proof of  \Cref{thm:full-info-adversarial}. 

\subsection{Proof of \Cref{lem:OGD-CB-entering-frequency}}

We demonstrate by analyzing the dynamic process of $\lambda_t$. Intuitively, the trajectory of $\lambda_t$ can be divided into consecutive phases, based on whether $\lambda_t = 0$ or $\lambda_t > 0$. 

Specifically, we prove by induction. When $t = 2$, the lemma naturally holds since \Cref{alg:OGD-CB} chooses to enter in the first round, and therefore $|\mathcal I_2| = 1$. 

Now suppose the lemma holds for all $2 \leq t \leq \tau$. When $t = \tau + 1$, we consider by cases on whether $\lambda_\tau = 0$. 

\paragraph{Case 1: $\lambda_\tau = 0$.} This case is rather trivial. Notice that by definition $\estr_\tau(v_\tau) \geq \epsilon_\tau v_\tau > 0$, therefore, $\estr_\tau(v_\tau) - \lambda_\tau \estc_\tau(v_\tau) > 0$ and $x_\tau = 1$ by the decision rule. As a result, 
\[|\mathcal I_{\tau + 1}| = |\mathcal I_{\tau}| + 1 \geq \constantFrequency(\tau - 1) + 1 \geq \constantFrequency\tau. \]

\paragraph{Case 2: $\lambda_\tau > 0$.} In this case, we define $l(\tau) := \max\{t < \tau: \lambda_t = 0\}$ be the last time $t < \tau$ such that $\lambda_t = 0$. Note that such $l(t)$ is well-defined since $\lambda_2 = 0$. We further define $\tau^+ := |\{l(\tau) \leq t < \tau: x_t = 1\}|$ be the rounds in $[l(\tau), \tau)$ such that \Cref{alg:OGD-CB} chooses to enter. We now give a lower bound on $\tau^+ / (\tau - l(\tau) + 1)$. 

Notice that by the definition of $l(\tau)$, we have $\lambda_{t + 1} - \lambda_t = \eta_t(x_t \estc_t(v_t) - \rho)$ for any $l(\tau) \leq t < \tau$. As a result, we have
\begin{align*}
    \lambda_\tau - \lambda_{l(\tau)} 
    &= \sum_{t = l(\tau)}^{\tau - 1} (\lambda_{t + 1} - \lambda_t) = \sum_{t = l(\tau)}^{\tau - 1} \eta_t(x_t \estc_t(v_t) - \rho) \\
    &\overset{(\text a)}{\leq} (\vmax - \rho)\sum_{t = l(\tau)}^{l(\tau) + \tau^+ - 1} \eta_t - \rho \sum_{t = l(\tau) + \tau^+}^{\tau - 1} \eta_t = \sum_{t = l(\tau)}^{l(\tau) + \tau^+ - 1} \eta_t - \rho \sum_{t = l(\tau)}^{\tau - 1} \eta_t. 
\end{align*}
Here, (a) holds since when $x_t = 1$, $x_t\estc_t(v_t) - \rho \leq \vmax - \rho$. Further, $\eta_t$ is decreasing in $t$. Now that we have $\lambda_\tau > 0 = \lambda_{l(\tau)}$. Meanwhile, for any $2\leq t_1 \leq t_2$, we have 
\[2(\sqrt{t_2 + 1} - \sqrt{t_1}) \leq \sum_{t = t_1}^{t_2} 1/\sqrt{t} \leq 2(\sqrt{t_2} - \sqrt{t_1 - 1}). \]
Therefore, noticing that $\eta_t = 1 / (\vmax \sqrt{t})$, we have
\begin{align*}
    0 &\leq \lambda_{\tau} - \lambda_{l(\tau)} \leq (\vmax - \rho)\sum_{t = l(\tau)}^{l(\tau) + \tau^+ - 1} \eta_t - \rho \sum_{t = l(\tau) + \tau^+}^{\tau - 1} \eta_t = \vmax \sum_{t = l(\tau)}^{l(\tau) + \tau^+ - 1} \eta_t - \rho \sum_{t = l(\tau)}^{\tau - 1} \eta_t \\
    &\leq (2 / \vmax)\cdot \left(\vmax\cdot\left(\sqrt{l(\tau) + \tau^+ - 1} - \sqrt{l(\tau) - 1}\right) - \rho\cdot \left(\sqrt{\tau} - \sqrt{l(\tau)}\right)\right). 
\end{align*}
That is, $\sqrt{l(\tau) + \tau^+ - 1} - \sqrt{l(\tau) - 1} \geq (\rho / \vmax)\cdot (\sqrt{\tau} - \sqrt{l(\tau)})$. We temporarily write $u := \sqrt{\tau} - \sqrt{l(\tau)}$ for simplicity, and consequently derive that
\begin{align*}
    \frac{\tau^+}{\tau - l(\tau) + 1} 
    &\geq \frac{\left((\rho / \vmax) u + \sqrt{l(\tau) - 1}\right)^2 - (l(\tau) - 1)}{\left(u + \sqrt{l(\tau)}\right)^2 - l(\tau)} \cdot \frac{\tau - l(\tau)}{\tau - l(\tau) + 1} \\
    &\overset{(\text a)}{\geq} \frac{1}{2}\cdot \frac{(\rho / \vmax)^2 u^2 + 2(\rho / \vmax) u \sqrt{l(\tau) - 1}}{u^2 + 2u\sqrt{l(\tau)}} 
    \overset{(\text b)}{\geq} \min\left\{\frac{\rho^2}{2\vmax^2}, \frac{\sqrt{2}\rho}{4\vmax}\right\} = \constantFrequency. 
\end{align*}
Here, (a) establishes as $\tau - l(\tau) \geq 1$, and (b) holds since $l(\tau) \geq 2$. Hence, we derive that
\[|\mathcal I_{\tau + 1}| \geq |\mathcal I_{l(\tau)}| + \tau^+ \geq \constantFrequency(l(\tau) - 1) + \constantFrequency(\tau - l(\tau) + 1) = \constantFrequency\tau. \]

By combining the two cases and induction, we conclude that \Cref{lem:OGD-CB-entering-frequency} holds. 

\subsection{Proof of \Cref{thm:partial-info-stochastic}}
According to \Cref{lem:OGD-CB-entering-frequency}, we have 
\[\epsilon_t = \sqrt{\frac{\log2 + 2\log T}{2|\mathcal I_t|}} \leq \sqrt{\frac{\log2 + 2\log T}{2C_e(t - 1)}}. \]
Hence, 
\begin{align*}
    \sum_{t = 2}^{T_0} \epsilon_t \leq \sum_{t = 2}^{T_0} \sqrt{\frac{\log2 + 2\log T}{2C_e(t - 1)}} \leq \sqrt{\frac{2T(\log 2 + 2\log T)}{C_e}}. 
\end{align*}
Meanwhile, the failure probability in estimation is also bounded: 
\begin{align*}
    \sum_{t = 2}^{T_0} q'(t, \epsilon_t) = 2\sum_{t = 2}^{T_0} \exp (-2|\mathcal I_t|\epsilon_t^2) \leq 1/T. 
\end{align*}

The remaining proof of \Cref{thm:partial-info-stochastic} is almost identical to the proof of \Cref{thm:full-info-stochastic}, and we omit it. 

\subsection{Proof of \Cref{thm:partial-info-adversarial}}

Similar with the proof of \Cref{thm:partial-info-stochastic}, according to \Cref{lem:OGD-CB-entering-frequency}, we adjust the construction of $\epsilon_t$ to $\sqrt{(\log2 + 2\log T)/(2|\mathcal I_t|)}$ in the proof of \Cref{thm:full-info-adversarial}. Since the proof is almost identical, we omit it.
\section{Proof of \Cref{thm:linear-difference-throttling-pacing}} \label{sec:proof-comparison}

To prove the theorem, we dig into the optimal solution structure of $\OPT$ and $\OPTstrong$, which is also widely studied previously. Recall that $r(v) \coloneqq \mathbb E_{p}[(v - p)^+]$ and $c(v) \coloneqq \mathbb E_{p}[p\mathbf 1[v \geq p]]$, we have the following proposition. 
\begin{proposition}[From \cite{talluri1998analysis}]\label{prop:optimal-solution-structure}
    When $\mathbb E_{v, p}[p \mathbf 1[v \geq p]] > \rho$, there exists two constants $\OPTthreshold, \OPTstrongThreshold \geq 0$ such that some optimal solution to $\OPT$ and $\OPTstrong$ has the following form, respectively:     
    \begin{itemize}
        \item For $\OPT$: $\pi(v) = 1$ if $r(v) > \OPTthreshold c(v)$, $\pi(v) = 0$ if $r(v) < \OPTthreshold c(v)$; 
        \item For $\OPTstrong$: $\kappa(v, p) = 1$ if $v > (1 + \OPTstrongThreshold) p$, $\kappa(v, p) = 0$ if $v < (1 + \OPTstrongThreshold) p$. 
    \end{itemize}
\end{proposition}

For a better view, we plot the result of \Cref{prop:optimal-solution-structure} in \Cref{fig:optimal-solution-structure}. We let $\OPTthreshold$ and $\OPTstrongThreshold$ be as defined in \Cref{prop:optimal-solution-structure}, and let
\begin{gather*}
    \Pi^- \coloneqq \{(v, p) \mid r(v) < \OPTthreshold c(v)\}, \quad \Pi^+ \coloneqq \{(v, p) \mid r(v) > \OPTthreshold c(v)\}; \\
    \Kappa^- \coloneqq \{(v, p) \mid v < (1 + \OPTstrongThreshold) p\}, \quad \Kappa^+ \coloneqq \{(v, p) \mid v > (1 + \OPTstrongThreshold) p\}. 
\end{gather*}
We now argue that as long as $\Pi^- \cap \Kappa^+$ has a positive measure concerning distributions $F$ and $G$, $\OPTstrong - \OPT = \Theta(T)$ holds when the budget constraint is strictly binding. The reason is that for the optimal solution $\pi^*$ of $\OPT$ specified in \Cref{prop:optimal-solution-structure}, we can take $\kappa(v, p) = 1$ for $(v, p) \in \Pi^- \cap \Kappa^+$, let $\kappa(v, p) = 0$ for some $(v, p) \in [0, \vmax]^2 \setminus \Pi^-$ with $v \leq (1 + \OPTstrongThreshold) p$, and keep $\kappa(v, p) = \pi^*(v)$ for other $(v, p)$ such that the budget constraint is still binding. This construction is possible by the feasibility of $\OPTstrong$ and the optimality of $\pi^*$. With such constructed $\kappa$, the objective value is larger than $\OPT$, and therefore $\OPTstrong / T > \OPT / T$. 

\begin{figure}
    \centering
    \begin{tikzpicture}
    \begin{axis}[
        xmin=0.25, xmax=3.25, 
        ymin=0, ymax=3.25,
        axis lines=middle,
        ticks=none,
        enlargelimits={abs=0.65},
        axis line style={-{Stealth[angle'=45]}},
        samples=100,
        domain=-1:3,
        clip=false
    ]
        
        \node[label={225:{$O$}}, circle, fill, inner sep=1pt] at (axis cs:0,0) {};
        \node[label={180:{$p$}}] at (axis cs:0,3) {};
        \node[label={270:{$v$}}] at (axis cs:3.25,0) {};

        \foreach \X in {0.6, 1.6, 2.5}
            \addplot[dashed, thick]
                coordinates {
                    (\X,-0.65) (\X,3+0.65)
                };

        \draw[pattern=north east lines, draw=none] (0,0) rectangle (0.6,1.4);
        \draw[pattern=north east lines, draw=none] (0,2.8+0.65-1.4) rectangle (0.6,2.8+0.65);
        \node at (0.3,1.725) {$>$};

        \draw[pattern=north east lines, draw=none] (1.6,0) rectangle (2.5,1.4);
        \draw[pattern=north east lines, draw=none] (1.6,2.8+0.65-1.4) rectangle (2.5,2.8+0.65);
        \node at (0.45+1.6,1.725) {$>$};

        \node at (1.1,1.725) {$<$};
        \node at (3.025,1.725) {$<$};

        \node at (1.75,4.35) (rv) {$r(v)= \OPTthreshold c(v)$};
        \node at (1.5,-1) {OPT};

        \draw[-{Stealth[angle'=45]}] (0.6,3+0.65) -- (1.15,4);
        \draw[-{Stealth[angle'=45]}] (1.6,3+0.65) -- (1.7,4);
        \draw[-{Stealth[angle'=45]}] (2.5,3+0.65) -- (2.25,4);
    \end{axis}

    \begin{scope}[shift={(8,0)}]
        \begin{axis}[
            xmin=0.25, xmax=3.25, 
            ymin=0, ymax=3.25,
            axis lines=middle,
            ticks=none,
            enlargelimits={abs=0.65},
            axis line style={-{Stealth[angle'=45]}},
            samples=100,
            domain=-1:3,
            clip=false
        ]
            
            \node[label={225:{$O$}}, circle, fill, inner sep=1pt] at (axis cs:0,0) {};
            \node[label={180:{$p$}}] at (axis cs:0,3) {};
            \node[label={270:{$v$}}] at (axis cs:3.25,0) {};
    
            \addplot[domain=0:3.85, dashed, thick] {0.9*x};
            \addplot[domain=0:3.65, pattern=north east lines, draw=none] {0.9*x} \closedcycle;
            \node at (1.5,-1) {$\OPTstrong$};
            \node at (1.75,4.35) (v) {$v=(1 + \OPTstrongThreshold)p$};
            \draw[-{Stealth[angle'=45]}] (3,3*0.9) to [bend left=15] (1.75,4);
        \end{axis}
    \end{scope}
\end{tikzpicture}
    \caption{Optimal solution structure of $\OPT$ and $\OPTstrong$ given in \Cref{prop:optimal-solution-structure}. }
    \label{fig:optimal-solution-structure}
\end{figure}

Similarly, when $\Pi^+ \cap \Kappa^-$ has a positive measure, we also have $\OPTstrong - \OPT = \Theta(T)$. Combining with the structure of $\Kappa^-$ and $\Kappa^+$, we have the following lemma: 
\begin{lemma}\label{lem:sufficient-condition-linear-difference}
    If there exists a constant $\delta > 0$ and a positive-measure set $V \subseteq [0, \vmax]$ with $V \times [0, \vmax] \subseteq \Pi^- \cup \Pi^+$, such that for each $v \in V$, 
    \begin{gather}\label{eq:G-continuity}
        \min \left\{\Pr_{p \sim G}\left[p > v / (1 + \OPTstrongThreshold)\right], \Pr_{p \sim G}\left[p < v / (1 + \OPTstrongThreshold)\right]\right\} \geq \delta, 
    \end{gather}
    then 
    \[\OPTstrong - \OPT = \Theta(T). \]
\end{lemma}

We now come back to the proof of the theorem. Let $\OPTthreshold$ and $\OPTstrongThreshold$ be defined as in \Cref{prop:optimal-solution-structure}. By \Cref{assump:r/c-non-constant}, the measure that $r(v) / c(v) \neq \OPTthreshold$ is positive, that is, the measure of $\Pi^- \cup \Pi^+$ is positive. This further implies that for some $\epsilon > 0$, $(\Pi^- \cup \Pi^+) \cap [\epsilon, 1 - \epsilon]$ has non-zero measure. Therefore, take $V$ be this set, and by \Cref{assump:G-continuity}, \eqref{eq:G-continuity} holds for some small constant $\delta > 0$. Applying \Cref{lem:sufficient-condition-linear-difference}, we have $\OPTstrong - \OPT = \Theta(T)$. Further synthesizing this with \Cref{prop:throttling-stochastic-lower-upper,,prop:pacing-stochastic-upper,,prop:hindsight-DLP-difference}, the proof is finished. 

\section{Missing Proofs in the Appendix}

\subsection{Proof of \Cref{lem:upper-bound-revenue-throttling}}
\label{app:upper-bound-revenue-throttling}
Given $\bm p$ including $S$ 1/3(s) and $(T - S)$ 2/3(s), we let $n_1 \leq S$ and $n_2 \leq T - S$ be the be the number of auctions in which the buyer enters with $p = 1/3$ and $p = 2/3$ respectively. Then by the budget constraint, we have
\[\frac{1}{3} \cdot n_1 + \frac{2}{3} \cdot n_2 \leq \frac{1}{2}\cdot T. \]
Meanwhile, the total revenue gained by $\beta$ under $\bm p$ is 
\[\totalRevenueNoV = \frac{2}{3} \cdot n_1 + \frac{1}{3} \cdot n_2. \] 

When $S \geq T/2$, then since $n_1 \leq S$ and $n_2 \leq T - S$, it is certain that 
\[\totalRevenueNoV \leq \frac{2}{3} \cdot S + \frac{1}{3} \cdot (T - S) = \frac{1}{3}\cdot (S + T). \]

When $S < T/2$, if $S$ is even, then
\begin{align*}
    \totalRevenueNoV &= \frac{2}{3} \cdot n_1 + \frac{1}{3} \cdot n_2 = \frac{1}{2} n_1 + \frac{1}{2} \left(\frac{1}{3} \cdot n_1 + \frac{2}{3} \cdot n_2\right) \\
    &\leq \frac{1}{2}S + \frac{1}{4}T, 
\end{align*}
and the equality holds with $n_1 = S$ and $n_2 = (3T - 2S) / 12$. On the other hand, if $S$ is odd, then by a similar argument the maximal value $\totalRevenueNoV$ can reach with $n_1 \leq S - 1$ is no larger than $(S - 1) / 2 + T/4$. If $n_1 = S$, then
\begin{align*}
    \totalRevenueNoV = \frac{2}{3} \cdot n_1 + \frac{1}{3} \cdot n_2 \leq \frac{2}{3} \cdot S + \frac{1}{3} \left\lfloor \frac{T / 2 - S / 3}{2 / 3} \right\rfloor = \frac{3T + 6S - 2}{12}. 
\end{align*}
Putting together, we have $\totalRevenueNoV \leq (3T + 6S - 2) / 12$. Combining both cases on whether $2|S$ or not, we derive that \[\totalRevenueNoV \leq \frac{2}{3} \cdot S + \frac{1}{3} \cdot  \left\lfloor \frac{3T - 2S}{4} \right\rfloor. \]

\subsection{Proof of \Cref{lem:regret-lower-bound-non-simplified}}
\label{app:regret-lower-bound-non-simplified}
We first calculate an upper bound on the expected total revenue of online throttling. Specifically, since $4|T$, we have
\begin{align*}
    \totalRevenueNoV &\overset{(\text a)}{=} \frac{1}{2^T}\left(\sum_{S = 0}^{T/2 - 1} \left(\frac{2}{3} S + \frac{1}{3} \left\lfloor \frac{3T - 2S}{4} \right\rfloor\right) \binom{T}{S} + \sum_{S = T/2}^{T} \frac{1}{3} (S + T) \binom{T}{S}\right) \\
    &\overset{(\text b)}{=} \frac{1}{2^T}\left(\sum_{S = 0}^{T/2 - 1} \left(\frac{2}{3} T + \frac{1}{3} S + \frac{1}{3} \left\lfloor \frac{3T - 2S}{4} \right\rfloor\right) \binom{T}{S} + \frac{T}{2} \binom{T}{T/2}\right). 
\end{align*}
Here, (a) is by \Cref{lem:upper-bound-revenue-throttling}, and (b) holds by noticing that $\binom{T}{S} = \binom{T}{T - S}$. 
Now notice that
\begin{align*}
    \OPT = \frac{T}{2} = \frac{1}{2^T}\sum_{S = 0}^T \frac{T}{2}\binom{T}{S} = \frac{1}{2^T} \left(\sum_{S = 0}^{T/2 - 1} T\binom{T}{S} + \frac{T}{2}\binom{T}{T/2}\right),
\end{align*}
and therefore, we derive that
\begin{align*}
    \OPT - \totalRevenueNoV &\geq \frac{1}{2^T} \sum_{S = 0}^{T/2 - 1} \frac{1}{3}\left(T - S - \left\lfloor \frac{3T - 2S}{4} \right\rfloor\right)\binom{T}{S} \\
    &\overset{(\text a)}{=} \frac{1}{3\cdot 2^T} \sum_{t = 1}^{T/4} \left(\frac{T}{4} - t + 1\right) \left(\binom{T}{2t - 2} + \binom{T}{2t - 1}\right) \\
    &\overset{(\text b)}{=} \frac{1}{12\cdot 2^T} \sum_{t = 1}^{T/4} \left(T - 4(t - 1)\right)\binom{T + 1}{2t - 1}. 
\end{align*}
Here, (a) holds by noticing that the multipliers of both $\binom{T}{2t - 2}$ and $\binom{T}{2t - 1}$ in the summing part equal to $T/4 - t + 1$. Meanwhile, (b) establishes since $\binom{T}{2t - 2} + \binom{T}{2t - 1} = \binom{T + 1}{2t - 1}$. 

\subsection{Proof of \Cref{lem:lower-bound-simplified}}
\label{app:lower-bound-simplified}
To start with, by binomial theorem, we have
\[\sum_{t = 1}^{T/2} \binom{T}{2t - 1} = \sum_{t = 1}^{T/2 + 1} \binom{T}{2t - 2} = 2^{T - 1}, \]
by adopting $\binom{T}{t} = \binom{T}{T - t}$ again, we obtain
\begin{gather}
    \sum_{t = 1}^{T/4} \binom{T}{2t - 1} = 2^{T - 2}, \quad
    \sum_{t = 1}^{T/4} \binom{T}{2t - 2} = 2^{T - 2} - \frac{1}{2}\binom{T}{T/2}.  \label{eq:two-binomials}
\end{gather}

With the help of these two equalities, we have the following computation: 
\begin{align*}
    \sum_{t = 1}^{T/4} \left(T - 4(t - 1)\right)\binom{T + 1}{2t - 1} &= (T + 2) \sum_{t = 1}^{T/4} \binom{T + 1}{2t - 1} - 2 \sum_{t = 1}^{T/4} (2t - 1)\binom{T + 1}{2t - 1} \\
    &\overset{(\text a)}{=} (T + 2) \sum_{t = 1}^{T/4} \left(\binom{T}{2t - 1} + \binom{T}{2t - 2}\right) - 2(T + 1) \sum_{t = 1}^{T/4} \binom{T}{2t - 2} \\
    &\overset{(\text b)}{=} (T + 2)\sum_{t = 1}^{T/4} \binom{T}{2t - 1} - T\sum_{t = 1}^{T/4} \binom{T}{2t - 2} \\
    &= 2^{T - 1} + \frac{T}{2}\binom{T}{T/2}. 
\end{align*}
For the above, (a) is by $(2t - 1)\binom{T + 1}{2t - 1} = (T + 1)\binom{T}{2t - 2}$, and (b) is derived by \eqref{eq:two-binomials}.

\subsection{Proof of \Cref{lem:yao-principle}}

Let $\Pr[\bm v^i]$ denote the probability of $\mathcal V$ choosing $\bm v^i$ for $i\in \{1, 2, \cdots m\}$. We then have the following sequence for any distribution $G'$:
\begin{align*}
    \inf_{\bm v, G}\mathbb{E}_{\bm p\sim G^T, \bm \gamma} \left[\discountedRegretGamma\right]
    &\overset{(\text a)}{=} \sum_{i=1}^m \Pr[\bm v^i] \inf_{\bm v, G}\mathbb{E}_{\bm p\sim G^T, \bm \gamma} \left[\discountedRegretGamma\right] \\
    &\overset{(\text b)}{\leq} \sum_{i=1}^m \Pr[\bm v^i] \mathbb{E}_{\bm p\sim (G')^T, \bm \gamma} \left[Reg_\mu^\beta(\bm v^i, \bm p, \bm \gamma)\right] \\
    &\overset{(\text c)}{=}  \mathbb{E}_{\bm p\sim (G')^T, \bm \gamma} \left[\sum_{i=1}^m \Pr[\bm v^i] Reg_\mu^\beta(\bm v^i, \bm p, \bm \gamma)\right] \\
    &\overset{(\text d)}{=} \mathbb{E}_{\bm p\sim (G')^T, \bm \gamma, \bm v\sim \mathcal{V}} \left[Reg_\mu^\beta(\bm v^i, \bm p, \bm \gamma)\right] \\
    &\leq \sup_{\bm \gamma}\mathbb{E}_{\bm p\sim (G')^T, \bm v \sim \mathcal V} \left[\discountedRegretGamma\right]. 
\end{align*}

In the above, (a) holds because of $\sum_{i=1}^m \Pr[\bm v^i] = 1$; (b) follows since $G'$ and $\bm v^i$ are feasible realizations of $G$ and $\bm v$, respectively; (c) is due to the additivity of expectation; At last, (d) follows from the independence of $G'$, $\gamma$, and $\mathcal V$.

\end{document}